\documentclass[12pt,english,floatfix,superscriptaddress,aps,prd,preprint,nofootinbib]{revtex4}
\usepackage{amsmath}
\usepackage{amssymb}
\usepackage{amsbsy}
\usepackage[a4paper, margin=1.6cm]{geometry}
\usepackage{amsfonts}
\usepackage{amsopn}
\usepackage{amstext}
\usepackage{graphicx}
\usepackage{amssymb}
\usepackage{amsfonts}
\usepackage{amsmath}
\usepackage{graphicx}
\usepackage[english]{babel}
\usepackage{color}
\usepackage{xcolor}
\usepackage{slashed}
\usepackage{esint}
\usepackage[dvips]{epsfig}
\usepackage[dvips]{graphicx}
\usepackage{float}
\usepackage{units}
\usepackage{textcomp}
\usepackage{placeins}
\usepackage{hyperref}             
\hypersetup{
    colorlinks=true,              
    breaklinks=true,              
    citecolor=blue,               
    linkcolor=[rgb]{0,0.5,0.9},   
    urlcolor=red,                 
    filecolor=green               
}

\usepackage{hyperref}
\usepackage{slashed}

\newcommand{\ie}{\begin{equation}}
\newcommand{\fe}{\end{equation}}
 \newcommand{\bq}{\begin{equation}}
 \newcommand{\eq}{\end{equation}}
 \newcommand{\bqn}{\begin{eqnarray}}
 \newcommand{\eqn}{\end{eqnarray}}

\begin{document}

\title{Gravitational wave propagation in Hořava–Lifshitz gravity}



\author{A. A. Ara\'{u}jo Filho}
\email{dilto@fisica.ufc.br}
\affiliation{Departamento de Física, Universidade Federal da Paraíba, Caixa Postal 5008, 58051--970, João Pessoa, Paraíba,  Brazil.}
\affiliation{Departamento de Física, Universidade Federal de Campina Grande Caixa Postal 10071, 58429-900 Campina Grande, Paraíba, Brazil.}
\affiliation{Center for Theoretical Physics, Khazar University, 41 Mehseti Street, Baku, AZ-1096, Azerbaijan.}
\author{J. L. A. Silva}
\email{joselucas21042@gmail.com}
\affiliation{Departamento de Física, Universidade Federal de Campina Grande Caixa Postal 10071, 58429-900 Campina Grande, Paraíba, Brazil.}


\author{N. Heidari}
\email{heidari.n@gmail.com}

\affiliation{Departamento de Física, Universidade Federal de Campina Grande Caixa Postal 10071, 58429-900 Campina Grande, Paraíba, Brazil.}
\affiliation{Center for Theoretical Physics, Khazar University, 41 Mehseti Street, Baku, AZ-1096, Azerbaijan.}
\affiliation{School of Physics, Damghan University, Damghan, 3671641167, Iran.}


\author{Jie Zhu}
\email{jiezhu@pku.edu.cn}

\affiliation{Department of Physics, Chongqing University, Chongqing 401331, P.R. China}


\author{Iarley P. Lobo}
\email{lobofisica@gmail.com}
\affiliation{Departamento de Física, Universidade Federal da Paraíba, Caixa Postal 5008, 58051--970, João Pessoa, Paraíba,  Brazil.}
\affiliation{Departamento de Física, Universidade Federal de Campina Grande Caixa Postal 10071, 58429-900 Campina Grande, Paraíba, Brazil.}


\author{V. B. Bezerra}
\email{valdir@fisica.ufpb.br}
\affiliation{Departamento de Física, Universidade Federal da Paraíba, Caixa Postal 5008, 58051--970, João Pessoa, Paraíba,  Brazil.}

\date{\today}

\begin{abstract}
We investigate the generation and propagation of gravitational waves in the leading parity--even infrared truncation of Hořava--Lifshitz gravity, characterized by the modified tensor dispersion relation $\omega^{2}=k^{2}+\alpha k^{4}$. Working in the transverse--traceless sector, we show that the higher--spatial--derivative correction preserves the conventional plus and cross polarizations and introduces neither polarization mixing, helicity splitting, nor gravitational birefringence. We construct the retarded Green function of the modified wave operator and derive the radiation--zone waveform to first order in $\alpha$. The resulting signal exhibits a frequency--dependent amplitude renormalization together with a dispersive propagation phase that accumulates over the source--observer distance. We apply the formalism to a binary black hole system in a quasi--circular orbit and obtain the polarization waveforms for an arbitrary observation direction. We further derive the corresponding energy flux, total luminosity, and adiabatic chirp evolution. In terms of the observed gravitational wave frequency $f$, the leading corrections satisfy $\Delta h_{A}/h_{A}^{\mathrm{GR}}=-8\pi^{2}\alpha f^{2}$ and $\Delta P/P_{\mathrm{GR}} =\Delta\dot{f}/\dot{f}_{\mathrm{GR}} =-16\pi^{2}\alpha f^{2}$, while the accumulated generation phase has the frequency dependence of a relative third post--Newtonian contribution. By mapping the Hořava--Lifshitz coefficient to the LIGO--Virgo--KAGRA modified--dispersion parametrization, we obtain $-6.2\times10^{2}\,\mathrm{eV}^{-2} <\alpha< 1.9\times10^{2}\,\mathrm{eV}^{-2}$ at $90\%$ credibility from the GWTC--4.0 posterior. 
\end{abstract}


\maketitle
\tableofcontents
    

\section{Introduction}

General relativity has provided an exceptionally successful description of gravitational phenomena over a wide range of scales. Nevertheless, its perturbative quantization based on the Einstein--Hilbert action is nonrenormalizable, and the inclusion of relativistic higher--curvature operators, although capable of improving the ultraviolet behavior, generally introduces additional poles associated with higher time derivatives and perturbative ghost degrees of freedom \cite{Stelle:1977}. Ho\v{r}ava gravity was proposed as a different route to ultraviolet completion: Lorentz symmetry is not imposed as a fundamental symmetry at arbitrarily high energies, spacetime is endowed with a preferred foliation, and time and space obey an anisotropic scaling characterized by a dynamical exponent $z$ \cite{Horava:2009uw,Horava:2009if}. In $3+1$ dimensions, the choice $z=3$ permits operators containing up to six spatial derivatives while keeping the equations of motion second order in time. This construction improves the ultraviolet behavior without introducing the standard Ostrogradsky instability associated with higher time derivatives.

Since the original proposal, several formulations of the theory have been developed and their consistency has been examined in detail. The projectable and nonprojectable versions differ in the allowed spacetime dependence of the lapse function and in the structure of their constraint sectors \cite{SotiriouVisserWeinfurtner:2009,Mukohyama:2010review, Sotiriou:2011review}. Early analyses identified potential scalar instabilities and strong--coupling problems \cite{Charmousis:2009,PapazoglouSotiriou:2010}, motivating the healthy nonprojectable extension and its low--energy description in terms of a hypersurface orthogonal Einstein--aether, or khronometric, theory \cite{BlasPujolasSibiryakov:2010,BlasPujolasSibiryakov:2011, Jacobson:2010}. Considerable progress has also been achieved in the quantum theory. Perturbative renormalizability has been established for projectable Ho\v{r}ava gravity \cite{Barvinsky:2016}, asymptotic freedom has been demonstrated in $2+1$ dimensions \cite{Barvinsky:2017}, and recent studies of the complete marginal coupling flow in $3+1$ dimensions have identified asymptotically free trajectories compatible with the phenomenologically relevant region of the kinetic coupling \cite{BarvinskyKurovSibiryakov:2024}. Current reviews provide detailed accounts of the remaining theoretical challenges, the low--energy parameter space, cosmology, black holes, universal horizons, and the status of the ultraviolet completion \cite{Wang:2017review,HerreroValea:2023}.

Perturbation theory provides the direct connection between the microscopic structure of Ho\v{r}ava gravity and potentially measurable phenomena. Linear perturbations around Minkowski or cosmological backgrounds separate into scalar, vector, and tensor sectors, whose dynamics depend on the formulation and on the operators retained in the gravitational potential. In the tensor sector, higher spatial derivatives generate modified dispersion relations containing even powers such as $k^{4}$ and $k^{6}$, while parity--violating operators constructed from the three--dimensional Cotton tensor may also produce helicity--dependent odd powers of the wave number \cite{Mukohyama:2009,TakahashiSoda:2009,Wang:2010tensor, WangWuZhaoZhu:2013}. These terms can modify the primordial tensor spectrum, produce frequency--dependent propagation, and, when parity is broken, generate chiral gravitational waves. At low energies, the relativistic contribution dominates, whereas the higher spatial derivative operators appear as controlled corrections whose importance increases with frequency.

The direct detection of gravitational waves has transformed such corrections from formal possibilities into testable effects. Beginning with GW150914 \cite{LIGO:2016GW150914}, compact binary signals have opened a strong--field and dynamical regime in which the propagation speed, polarization content, amplitude evolution, and phase of gravitational radiation can be compared with the predictions of general relativity. The joint observation of GW170817 and GRB~170817A imposed a particularly stringent constraint on the low--energy tensor propagation speed \cite{LIGO:2017GW170817,LIGO:2017MultiMessenger}. The observational sample has continued to grow through GWTC--5.0 \cite{LVK:2026GWTC5}, while the most recent catalog wide parameterized tests relevant to modified gravitational wave dispersion have been reported with GWTC--4.0 \cite{LVK:2026GWTC4TestsII}. Up to now, no compelling evidence for a departure from general relativity has been found, but the increasing number, distance, bandwidth, and signal--to--noise ratio of the detected events substantially improve the sensitivity to small dispersive corrections accumulated over astrophysical propagation distances.

On the other hand, a modified dispersion relation changes the relation among frequency, wave number, phase velocity, and group velocity \cite{md1,md2,md3,md4,md5,md6,md7,md8,md9,md10,md11,md12,md13,md14,md15,md16,md17,md18,md19}. Consequently, different Fourier components of a compact binary signal can accumulate different propagation phases before reaching the detector. General parameterizations of Lorentz--violating dispersion have been incorporated into gravitational wave waveforms and data analyses \cite{MirshekariYunesWill:2012,KosteleckyMewes:2016}. Subsequent studies have clarified the treatment of higher harmonics, lensing degeneracies, and the distinction between group velocity and particle velocity prescriptions \cite{EzquiagaEtAl:2022,BakaEtAl:2025}. Higher--spatial--derivative operators have also been constrained directly through compact binary observations \cite{GongEtAl:2022}, and recent analyses within linearized effective gravity have studied anisotropy, birefringence, polarization mixing, retarded propagation, and waveform deformations generated by Lorentz-- and diffeomorphism--violating operators \cite{WangYanZhuZhao:2025,AraujoFilhoHeidariLobo:2026,A1,A2,A3}.

Gravitational radiation in Ho\v{r}ava gravity has previously been investigated from complementary perspectives. In the low--energy theory, the radiation emitted by binary systems can differ from the general relativistic quadrupole prediction because of modified tensor propagation and the possible excitation of an additional scalar degree of freedom \cite{BlasSanctuary:2011}. The complete polarization content of the theory and its compatibility with the GW170817 speed bound have also been studied through gauge invariant perturbations \cite{GongHouPapantonopoulosTzortzis:2018}. It is important, however, to distinguish the polarization content of the full theory from that of a restricted transverse--traceless tensor truncation. The additional scalar polarization belongs to the scalar sector, whereas an isotropic and parity--even correction acting diagonally on the transverse--traceless components does not, by itself, generate a new tensor polarization or split the two tensor helicities.

In this work, we investigate the generation and propagation of gravitational waves in the leading parity--even infrared truncation of Hořava--Lifshitz gravity. Restricting the analysis to the transverse--traceless tensor sector, we examine the polarization content associated with the modified dispersion relation $\omega^{2}=k^{2}+\alpha k^{4}$ and construct the corresponding retarded Green function. We then derive the radiation--zone waveform to first order in $\alpha$ and apply the resulting formalism to a binary black hole system in a quasi circular orbit. The associated polarization waveforms, energy flux, luminosity, and adiabatic chirp evolution are obtained, together with the leading amplitude and phase corrections. Finally, we map the Hořava--Lifshitz parameter $\alpha$ onto the modified dispersion framework employed by the LIGO--Virgo--KAGRA Collaboration and derive the corresponding observational constraint.


\section{A brief overview: Ho\v{r}ava gravity and tensor dispersion}

Ho\v{r}ava gravity is a quantum gravity motivated modification of general relativity based on an anisotropic scaling between time and space,
\begin{equation}
t \to b^{z}t, \qquad x^{i} \to b\,x^{i},
\end{equation}
where the critical exponent $z=3$ in $3+1$ dimensions leads to power counting renormalizability \cite{Horava:2009uw,Horava:2009if}. The theory is naturally formulated in terms of the Arnowitt--Deser--Misner (ADM) decomposition,
\begin{equation}
\mathrm{d}s^{2} = -N^{2}\mathrm{d}t^{2} + g_{ij} \left(\mathrm{d}x^{i}+N^{i}\mathrm{d}t\right) \left(\mathrm{d}x^{j}+N^{j}\mathrm{d}t\right),
\end{equation}
and is invariant under foliation preserving diffeomorphisms,
\begin{equation}
t \to t'(t), \qquad x^{i} \to x'^{i}(t,\vec{x}).
\end{equation}
This reduced symmetry permits the inclusion of higher--order spatial derivatives while keeping the equations of motion at most second order in time. It therefore avoids the Ostrogradsky instabilities associated with higher--time derivative operators, although additional stability conditions must still be imposed on the different propagating sectors of the theory.

The gravitational action can be written schematically as
\begin{equation}
S = \frac{M_{\rm Pl}^{2}}{2} \int \mathrm{d}t\,\mathrm{d}^{3}x\, N\sqrt{g} \left[ K_{ij}K^{ij} -\lambda K^{2} -\mathcal{V} \right],
\label{HL_action_general}
\end{equation}
where
\begin{equation}
K_{ij} = \frac{1}{2N} \left( \dot{g}_{ij} -\nabla_{i}N_{j} -\nabla_{j}N_{i} \right), \qquad K = g^{ij}K_{ij},
\end{equation}
is the extrinsic curvature of the preferred spatial slices. The potential $\mathcal{V}$ is constructed from three--dimensional curvature invariants and, in the nonprojectable theory, may also depend on $a_i \equiv \nabla_i \ln N$ and its spatial derivatives. At low energies, the potential contains the spatial curvature term required to recover the tensor sector of general relativity, whereas higher spatial derivative operators become increasingly relevant at high energies. A representative parity--even potential is
\begin{align}
\mathcal{V} ={}& -\xi R +\frac{1}{M_{\star}^{2}} \left( g_{2}R^{2} +g_{3}R_{ij}R^{ij} \right) \nonumber\\ &+ \frac{1}{M_{\star}^{4}} \left( g_{4}R\nabla^{2}R +g_{5}R_{ij}\nabla^{2}R^{ij} +g_{6}R^{3} +g_{7}R R_{ij}R^{ij} +g_{8}R^{i}{}_{j}R^{j}{}_{k}R^{k}{}_{i} +\cdots \right),
\label{HL_potential_generic}
\end{align}
where $M_{\star}$ denotes the characteristic scale suppressing the higher--curvature operators. In the original detailed balance construction, parity--odd terms involving the three--dimensional Cotton tensor may also be present. Such terms distinguish the two tensor helicities and introduce odd powers of the wave number into the corresponding dispersion relations.

To derive the tensor dispersion relation, we consider perturbations around Minkowski spacetime,
\begin{equation}
N = 1+\delta N, \qquad N_i = n_i, \qquad g_{ij} = \delta_{ij}+h_{ij},
\end{equation}
and isolate the transverse--traceless tensor sector, $\partial_i h_{ij} = 0$, and $ h^{i}{}_{i} = 0$. At linear order, $h_{ij}$ contains the two physical spin--2 polarizations, whereas the lapse and shift perturbations are nondynamical or belong to the scalar and vector sectors. After projecting onto the transverse--traceless sector, the linearized extrinsic curvature becomes
\begin{equation}
K_{ij}^{(1)} = \frac{1}{2}\dot{h}_{ij}, \qquad K^{(1)} = 0.
\end{equation}
As it is straightforward to verify, the coupling $\lambda$ does not affect tensor propagation at this order.

The linearized three--dimensional Ricci tensor and Ricci scalar are
\begin{equation}
R_{ij}^{(1)} = -\frac{1}{2}\nabla^{2}h_{ij}, \qquad R^{(1)} = 0.
\label{linear_ricci_TT}
\end{equation}
These expressions determine which operators in \eqref{HL_potential_generic} contribute to the linear tensor equation. Although $R^{(1)}=0$, the second--order expansion of the spatial Ricci scalar produces the standard two--spatial--derivative contribution. The operator $R_{ij}R^{ij}$ generates four spatial derivatives, whereas $R_{ij}\nabla^{2}R^{ij}$ generates six spatial derivatives. By contrast, $R^{2}$ and $R\nabla^{2}R$ do not contribute to the quadratic transverse--traceless action around Minkowski spacetime because $R^{(1)}=0$. The cubic--curvature operators begin at third order in the perturbations and therefore do not modify the linear tensor propagation.

The resulting quadratic action for the tensor perturbations can be written
schematically as
\begin{align}
S_{\rm T}^{(2)} ={}& \frac{M_{\rm Pl}^{2}}{8} \int \mathrm{d}t\,\mathrm{d}^{3}x \bigg[ \dot{h}_{ij}\dot{h}_{ij} -c_{T}^{2}\, \partial_{\ell}h_{ij}\partial_{\ell}h_{ij} -\frac{\gamma_{4}}{M_{\star}^{2}} \left(\nabla^{2}h_{ij}\right)^{2} \nonumber\\ &\hspace{3.5cm} -\frac{\gamma_{6}}{M_{\star}^{4}} \left(\partial_{\ell}\nabla^{2}h_{ij}\right) \left(\partial_{\ell}\nabla^{2}h_{ij}\right) +\cdots \bigg],
\label{quadratic_tensor_action_HL}
\end{align}
where $c_T$ is the low--energy tensor propagation speed, while $\gamma_4$ and $\gamma_6$ are dimensionless combinations of the couplings entering the potential. With the normalization adopted in \eqref{HL_action_general}, we may identify $c_T^{2}=\xi$ in the tensor sector. The relativistic normalization corresponds to $c_T=1$.

Varying \eqref{quadratic_tensor_action_HL} with respect to $h_{ij}$ gives
\begin{equation}
\ddot{h}_{ij} -c_T^{2}\nabla^{2}h_{ij} +\frac{\gamma_{4}}{M_{\star}^{2}}\nabla^{4}h_{ij} -\frac{\gamma_{6}}{M_{\star}^{4}}\nabla^{6}h_{ij} +\cdots =0.
\label{tensor_eom_HL}
\end{equation}
The relative minus sign multiplying $\nabla^{6}h_{ij}$ is required to produce a positive $\gamma_{6}k^{6}$ contribution to the dispersion relation.

We now consider plane wave solutions of the form
\begin{equation}
h_{ij}(t,\vec{x}) = \epsilon_{ij}\, e^{-i\omega t+i\vec{k}\cdot\vec{x}}, \qquad k \equiv |\vec{k}|,
\end{equation}
with $k_i\epsilon_{ij} = 0$, and $\epsilon^{i}{}_{i} = 0$. For these modes, we have
\begin{equation}
\nabla^{2}h_{ij} = -k^{2}h_{ij}, \qquad \nabla^{4}h_{ij} = k^{4}h_{ij}, \qquad \nabla^{6}h_{ij} = -k^{6}h_{ij}.
\end{equation}
Equation \eqref{tensor_eom_HL} therefore yields the parity--even tensor dispersion relation
\begin{equation}
\omega^{2} = c_T^{2}k^{2} +\frac{\gamma_{4}}{M_{\star}^{2}}\,k^{4} +\frac{\gamma_{6}}{M_{\star}^{4}}\,k^{6} +\cdots.
\label{HL_dispersion_tensor}
\end{equation}

In parity--violating formulations, including the original detailed balance construction, Cotton tensor contributions may additionally generate a helicity--dependent term containing five spatial derivatives. The two circular tensor polarizations then obey
\begin{equation}
\omega_{\pm}^{2} = c_T^{2}k^{2} +\frac{\gamma_{4}}{M_{\star}^{2}}\,k^{4} \pm \frac{\gamma_{5}}{M_{\star}^{3}}\,k^{5} +\frac{\gamma_{6}}{M_{\star}^{4}}\,k^{6} +\cdots,
\label{HL_dispersion_tensor_helicity}
\end{equation}
where the signs $\pm$ label the two graviton helicities. The $k^{5}$ contribution is absent when parity is preserved.

In the infrared regime, $k\ll M_{\star}$, the higher--order contributions are suppressed and the tensor modes recover an approximately relativistic dispersion relation,
\begin{equation}
\omega^{2} \simeq c_T^{2}k^{2}.
\end{equation}
When the six--derivative contribution dominates at high wave number, one instead obtains
\begin{equation}
\omega^{2} \simeq \frac{\gamma_{6}}{M_{\star}^{4}}\,k^{6} \propto k^{2z}, \qquad z=3,
\end{equation}
which displays the characteristic Lifshitz scaling of Ho\v{r}ava gravity.

For the purposes of the present work, we restrict the analysis to the leading parity--even correction beyond the relativistic term,
\begin{equation}
\omega^{2} = c_T^{2}k^{2}+\alpha k^{4}, \qquad \alpha \equiv \frac{\gamma_{4}}{M_{\star}^{2}}.
\label{HL_dispersion_truncated}
\end{equation}
The corresponding tensor equation is
$\left( \partial_t^{2} -c_T^{2}\nabla^{2} +\alpha\nabla^{4} \right)h_{ij} =0$.
After normalizing the low--energy tensor speed to unity, this equation reduces to
\begin{equation}
\left( \partial_t^{2} -\nabla^{2} +\alpha\nabla^{4} \right)h_{ij} =0.
\end{equation}
The truncated relation \eqref{HL_dispersion_truncated} should be understood as an effective description valid in the regime where the $k^{4}$ correction is relevant but the $k^{5}$ and $k^{6}$ contributions remain negligible, as we should expect. In other words, it captures the leading dispersive modification of the tensor modes but is not intended to describe the asymptotic $z=3$ ultraviolet regime. The effects associated with the helicity--dependent $k^{5}$ term and the six--derivative contribution will be considered in future investigations.


\section{Tensor polarizations in Ho\v{r}ava-Lifshitz gravity}

The polarization content of gravitational waves provides direct information  about the dynamical structure of the underlying gravitational theory. In the  Hořava--Lifshitz truncation considered here, the modifications arise from  higher--spatial--derivative operators that preserve spatial rotational  invariance. In this manner, the tensor polarization basis remains unchanged,  whereas the propagation of the corresponding modes becomes dispersive.

We work in the weak--field regime around the Minkowski background, $\eta_{\mu\nu}=\mathrm{diag}(-1,1,1,1)$. Restricting the analysis to the transverse--traceless tensor sector, the  spatial perturbations $h_{ij}(x)$ satisfy the modified linearized equation
\begin{equation}
\label{EOM_HL_TT}
\left(\partial_t^2-\nabla^2+\alpha\nabla^4\right)h_{ij}(x)=0,
\end{equation}
where $\alpha$ parametrizes the leading higher spatial derivative correction, with
$\nabla^2=\delta^{kl}\partial_k\partial_l$, and $\nabla^4=\left(\nabla^2\right)^2$.
Since $\nabla^4$ is rotationally invariant and contains no tensorial  structure capable of mixing different spatial directions, it acts identically  on all transverse--traceless components. In other words, it neither generates  additional tensor polarizations nor distinguishes between the existing ones.

The transverse--traceless conditions are
$h_{ii}=0$, and $\partial_i h_{ij}=0$. Within the tensor sector, these conditions leave two independent propagating  degrees of freedom. This statement applies specifically to the tensorial  perturbations considered here and does not, by itself, exclude additional  scalar modes that may arise in more general formulations of  Hořava--Lifshitz gravity, i.e., higher--order corrections.

To determine the polarization structure explicitly, we consider a plane wave  propagating along the $z$ direction,
\begin{equation}
h_{ij}(t,\mathbf{x}) = \varepsilon_{ij}\,e^{i(kz-\omega t)},
\end{equation}
where $k$ denotes the spatial wave number. Equivalently, the corresponding  wave four vector may be written as
$k^\mu=(\omega,0,0,k)$.
Substitution of the plane wave solution into Eq.~\eqref{EOM_HL_TT} yields the  modified dispersion relation
\begin{equation}
\omega^2=k^2+\alpha k^4.
\end{equation}

The spatial transversality condition,
$k_i\varepsilon_{ij}=0$, together with tracelessness,
$\delta^{ij}\varepsilon_{ij}=0$,
restricts the polarization tensor to the plane orthogonal to the direction of  propagation. In four--dimensional notation, with $\varepsilon_{0\mu}=0$ in the  tensor sector, the polarization tensor takes the form
\begin{equation}
\varepsilon_{\mu\nu}
=
\begin{pmatrix}
0 & 0 & 0 & 0 \\
0 & \varepsilon_{+} & \varepsilon_{\times} & 0 \\
0 & \varepsilon_{\times} & -\varepsilon_{+} & 0 \\
0 & 0 & 0 & 0
\end{pmatrix}.
\end{equation}
The independent amplitudes $\varepsilon_{+}$ and $\varepsilon_{\times}$  correspond to the conventional $+$ and $\times$ tensor polarizations.

Because the modified wave operator is isotropic, parity even, and diagonal in  polarization space, both tensor modes obey the same dispersion relation. Accordingly, this truncation produces neither helicity splitting nor polarization mixing or gravitational birefringence, as argued before.

\begin{figure}
    \centering
    \includegraphics[scale=0.5]{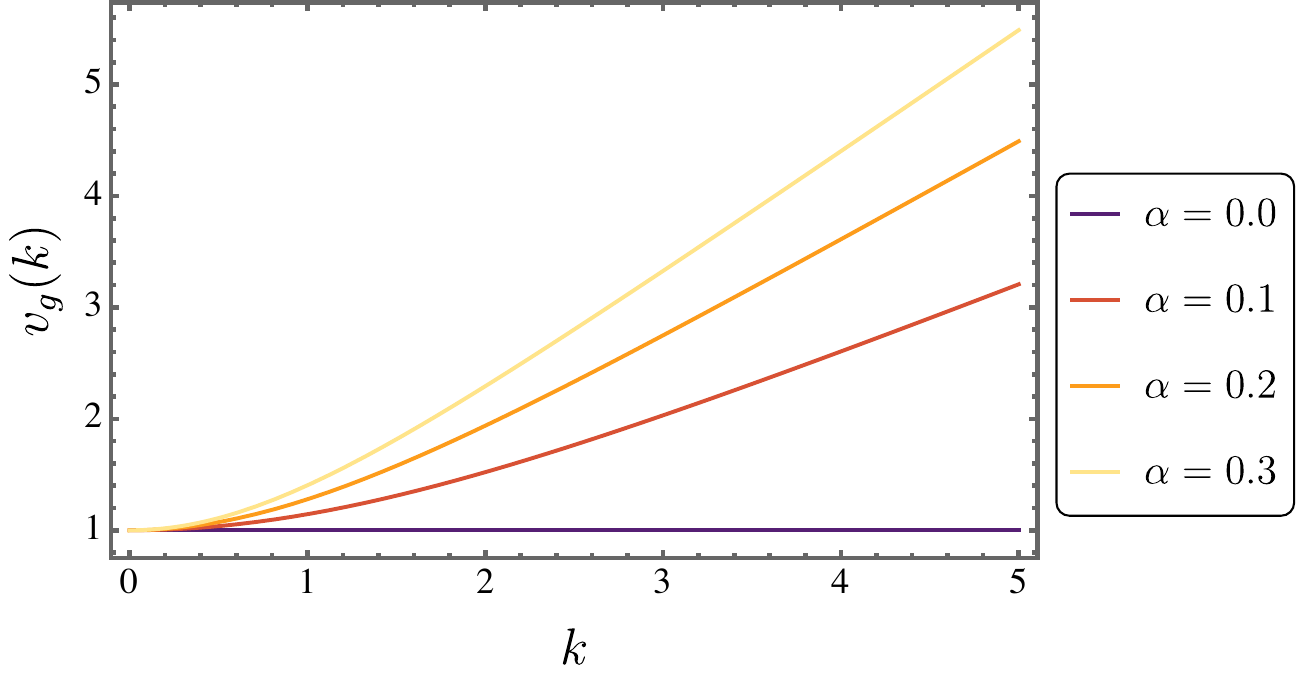}
    \caption{Group velocity $v_g$ as a function of the wave number $k$ for different values of the higher spatial derivative parameter $\alpha$. For $\alpha=0$, the standard relativistic result $v_g=1$ is recovered, whereas positive values of $\alpha$ lead to a wave number dependent group velocity that increases with both $k$ and $\alpha$.}
    \label{groupvelocity}
\end{figure}

Although the polarization content remains unchanged, the propagation is  modified. For the positive frequency branch,
$\omega(k)=\sqrt{k^2+\alpha k^4}$, the group velocity is
\begin{equation}
v_g = \frac{\mathrm{d}\omega}{\mathrm{d}k} = \frac{k+2\alpha k^3}{\sqrt{k^2+\alpha k^4}} = \frac{1+2\alpha k^2}{\sqrt{1+\alpha k^2}},
\end{equation}
where the last expression assumes $k>0$. The dependence of $v_g$ on the wave  number implies frequency--dependent propagation, leading to dispersive spreading and modifications of the phase evolution and temporal profile of  the gravitational wave signal. In Fig.~\ref{groupvelocity}, we present the group velocity $v_g$ as a function of the wave number $k$ for different values of the parameter $\alpha$. In the general relativistic limit, $\alpha=0$, the dispersion relation reduces to $\omega^{2}=k^{2}$, and the tensor modes propagate with the constant velocity $v_g=1$. As it is straightforward to see (and argued before), for $\alpha>0$, the higher spatial derivative contribution makes the propagation dispersive, with the group velocity depending explicitly on the wave number. The curves approach the relativistic value in the low--momentum regime, $k\rightarrow 0$, while increasingly departing from it as $k$ grows. Moreover, larger values of $\alpha$ enhance this deviation.

Notice that, in other words, within the isotropic $k^4$ Hořava--Lifshitz truncation, the two  transverse tensor polarizations retain the same structure as in general  relativity, while the higher--spatial--derivative correction modifies their  dispersion relation and propagation speed. The absence of anisotropic or  parity--odd operators ensures that the two polarizations remain dynamically  degenerate.


\section{Matter-sourced gravitational waves in Ho\v{r}ava-Lifshitz gravity}

\subsection{Retarded Green function and  propagation}

To describe gravitational wave generation, we extend the homogeneous equation for the transverse--traceless metric perturbation by coupling it to the transverse--traceless part of the matter stress tensor. In the normalization adopted throughout this work, the sourced field equation reads
\begin{equation}
\label{EOM_HL_source}
\left( \partial_t^2-\nabla^2+\alpha\,\nabla^4 \right) h_{ij}^{\mathrm{TT}}(x) = 16\pi G\,T_{ij}^{\mathrm{TT}}(x) \equiv J_{ij}^{\mathrm{TT}}(x).
\end{equation}
Since $\nabla^4$ contains two additional powers of inverse length relative to $\nabla^2$, dimensional consistency requires $[\alpha]=L^2$, and $(c=1)$. Accordingly, the perturbative regime is controlled not by $|\alpha|\ll 1$, which is not dimensionally meaningful, but by the dimensionless condition $|\alpha|\,k^2\ll 1$, and $k\equiv |\vec p|$. Equivalently, one may introduce the characteristic scale $\Lambda_{\alpha}=|\alpha|^{-1/2}$ and write $\alpha=s_{\alpha}\Lambda_{\alpha}^{-2}$, $s_{\alpha}\equiv \operatorname{sgn}(\alpha)$, $k\ll\Lambda_{\alpha}$. For $\alpha<0$, the truncated dispersion relation becomes unstable when $k>|\alpha|^{-1/2}$.

As shown below, the corresponding dispersion relation is
\begin{equation}
\omega^2=k^2+\alpha k^4.
\end{equation}
Thereby, in the perturbative regime $|\alpha|k^2\ll1$, the frequency and group velocity are given by
\begin{equation}
\omega = k\sqrt{1+\alpha k^2} \simeq k\left(1+\frac{\alpha k^2}{2}\right), \qquad v_g = \frac{\mathrm{d}\omega}{\mathrm{d}k} = \frac{1+2\alpha k^2}{\sqrt{1+\alpha k^2}} \simeq 1+\frac{3}{2}\alpha k^2.
\end{equation}
Notice that within the validity domain of the truncated theory, positive and negative values of $\alpha$ respectively increase and decrease the group velocity relative to its relativistic value. This statement concerns the propagation of wave packets and should not, by itself, be interpreted as a criterion for causal propagation.

The solution for the metric perturbation is obtained by convolving the source with the retarded Green function of the modified wave operator,
\begin{equation}
\label{hij_solution_HL}
h_{ij}^{\mathrm{TT}}(x) = \int \mathrm{d}^4y\, G^{\mathrm{ret}}(x-y)\, J_{ij}^{\mathrm{TT}}(y),
\end{equation}
where $G^{\mathrm{ret}}$ satisfies
\begin{equation}
\left( \partial_t^2-\nabla^2+\alpha\nabla^4 \right) G^{\mathrm{ret}}(t,\vec x) = \delta(t)\delta^{(3)}(\vec x), \qquad G^{\mathrm{ret}}(t>0,\vec x)=0.
\end{equation}

For a plane wave mode with spatial momentum $k=|\vec p|$, we introduce
\begin{equation}
\Omega_k \equiv \sqrt{k^2+\alpha k^4} = k\sqrt{1+\alpha k^2},
\end{equation}
which is real for modes satisfying $1+\alpha k^2>0$. In momentum space, the retarded Green function is
\begin{equation}
\widetilde G^{\mathrm{ret}}(\omega,k) = \frac{1}{\Omega_k^2-(\omega+i\epsilon)^2}.
\end{equation}
Equivalently, it can be written in factorized form as
\begin{equation}
\widetilde G^{\mathrm{ret}}(\omega,k) = \frac{1} {\bigl(\Omega_k-\omega-i\epsilon\bigr)  \bigl(\Omega_k+\omega+i\epsilon\bigr)} = -\frac{1} {\bigl(\omega-\Omega_k+i\epsilon\bigr)  \bigl(\omega+\Omega_k+i\epsilon\bigr)}.
\end{equation}
The retarded prescription places both poles in the lower half of the complex $\omega$ plane:
\begin{equation}
\omega_1=\Omega_k-i\epsilon, \qquad \omega_2=-\Omega_k-i\epsilon.
\end{equation}

The mixed time--momentum representation is defined by
\begin{equation}
\widetilde G^{\mathrm{ret}}(t,k) = \int_{-\infty}^{\infty} \frac{\mathrm{d}\omega}{2\pi}\, e^{-i\omega t}\, \widetilde G^{\mathrm{ret}}(\omega,k).
\end{equation}
For $t>0$, the integration contour is closed in the lower half plane, and both poles contribute. Evaluating the corresponding residues gives
\begin{equation}
\label{retarrrd}
\widetilde G^{\mathrm{ret}}(t,k) = \Theta(t)\, \frac{\sin(\Omega_k t)}{\Omega_k}.
\end{equation}

We now construct the perturbative derivative expansion of Eq.~\eqref{retarrrd}. To first order in $\alpha$, the modified frequency is
\begin{equation}
\Omega_k = k+\frac{\alpha k^3}{2} +\mathcal{O}(\alpha^2).
\end{equation}
In this manner,
\begin{equation}
\frac{\sin(\Omega_k t)}{\Omega_k} = \frac{\sin(kt)}{k} + \frac{\alpha}{2} \left[ k^2t\cos(kt)-k\sin(kt) \right] + \mathcal{O}(\alpha^2),
\end{equation}
and the mixed Green function becomes
\begin{equation}
\label{mixed_expand_HL}
\widetilde G^{\mathrm{ret}}(t,k) = \Theta(t) \left\{ \frac{\sin(kt)}{k} + \frac{\alpha}{2} \left[ k^2t\cos(kt)-k\sin(kt) \right] \right\} + \mathcal{O}(\alpha^2).
\end{equation}
In addition to $|\alpha|k^2\ll1$, the direct expansion of the accumulated oscillatory phase requires $|\alpha|k^3t\ll1$. This additional condition is particularly relevant for propagation over large distances. When it is not satisfied, the exact frequency $\Omega_k$ should be retained rather than expanded inside the oscillatory functions.

The coordinate space kernel follows from the inverse spatial Fourier transformation. For an isotropic function of $k$, we obtain \begin{equation}
\label{sffss}
G^{\mathrm{ret}}(t,r) = \frac{1}{2\pi^2r} \int_0^\infty \mathrm{d}k\, k\sin(kr)\, \widetilde G^{\mathrm{ret}}(t,k), \qquad r\equiv |\vec x|.
\end{equation}
Substituting Eq.~\eqref{retarrrd} into Eq.~\eqref{sffss} gives the exact radial representation
\begin{equation}
\label{Gret_radial_exact}
G^{\mathrm{ret}}(t,r) = \frac{\Theta(t)}{2\pi^2r} \int_0^\infty \mathrm{d}k\, k\sin(kr)\, \frac{\sin(\Omega_k t)}{\Omega_k}.
\end{equation}

Furthermore, using the perturbative result in Eq.~\eqref{mixed_expand_HL}, we get
\begin{align}
G^{\mathrm{ret}}(t,r) ={}& \frac{\Theta(t)}{2\pi^2r} \int_0^\infty \mathrm{d}k\, \sin(kr)\sin(kt) \nonumber\\ &+ \frac{\alpha\,\Theta(t)}{4\pi^2r} \int_0^\infty \mathrm{d}k\, \left[ k^3t\sin(kr)\cos(kt) - k^2\sin(kr)\sin(kt) \right] + \mathcal{O}(\alpha^2).
\end{align}
The momentum integrals are understood in the sense of tempered distributions. In order to step forward, we have to take into account some identities, which are
\begin{align}
\int_0^\infty \mathrm{d}k\, \sin(kr)\sin(kt) &= \frac{\pi}{2} \left[ \delta(t-r)-\delta(t+r) \right], \\[4pt] \int_0^\infty \mathrm{d}k\, k^2\sin(kr)\sin(kt) &= -\frac{\pi}{2} \left[ \delta''(t-r)-\delta''(t+r) \right], \\[4pt] \int_0^\infty \mathrm{d}k\, k^3\sin(kr)\cos(kt) &= -\frac{\pi}{2} \left[ \delta'''(t-r)-\delta'''(t+r) \right].
\end{align}
It follows that
\begin{align}
G^{\mathrm{ret}}(t,r) ={}& \frac{\Theta(t)}{4\pi r} \left[ \delta(t-r)-\delta(t+r) \right] \nonumber\\ &+ \frac{\alpha\,\Theta(t)}{8\pi r} \left\{ \delta''(t-r)-\delta''(t+r) - t\left[ \delta'''(t-r)-\delta'''(t+r) \right] \right\} + \mathcal{O}(\alpha^2).
\end{align}
In the physical region $t>0$ and $r>0$, the terms supported at $t=-r$ do not contribute. The retarded Green function therefore reduces to a more compact form as follows:
\begin{equation}
\label{Gret_final_HL}
G^{\mathrm{ret}}(t,r) = \Theta(t) \left[ \frac{\delta(t-r)}{4\pi r} + \frac{\alpha}{8\pi r} \left( \delta''(t-r)-t\,\delta'''(t-r) \right) \right] + \mathcal{O}(\alpha^2).
\end{equation}
Using the distributional identity
\begin{equation}
(t-r)\delta'''(t-r)=-3\delta''(t-r),
\end{equation}
Eq.~\eqref{Gret_final_HL} may equivalently be expressed as
\begin{equation}
G^{\mathrm{ret}}(t,r) = \Theta(t) \left[ \frac{\delta(t-r)}{4\pi r} + \frac{\alpha}{8\pi r} \left( 4\delta''(t-r)-r\,\delta'''(t-r) \right) \right] + \mathcal{O}(\alpha^2).
\label{greenfunddasdsa}
\end{equation}

The result above represents a local derivative expansion of the Green function. At any finite order in $\alpha$, this expansion is written in terms of derivatives of $\delta(t-r)$ and therefore remains supported on the general relativistic null cone. The retarded time appearing in this local representation is consequently $t_r=t-r$. Notice that the kernel in Eq.~\eqref{Gret_radial_exact} retains the frequency dependence encoded in $\Omega_k$ and generally develops support away from $t=r$.


\subsection{Radiation zone waveform}

Inserting the local Green function in Eq.~(\ref{greenfunddasdsa}) into the retarded solution, we have
\begin{equation}
h_{ij}^{\mathrm{TT}}(t,\mathbf r) = 16\pi G \int_{-\infty}^{+\infty}\mathrm{d}t' \int\mathrm{d}^{3}y\; G^{\mathrm{ret}}(t-t',R)\, T_{ij}^{\mathrm{TT}}(t',\mathbf y), \qquad R=|\mathbf r-\mathbf y|.
\label{retarded_solution_hij}
\end{equation}
Since the observer lies outside the compact source, $R>0$. The distributions entering the Green function are supported at $t-t'=R>0$, and therefore the Heaviside function is equal to unity on their support. Equation~\eqref{retarded_solution_hij} then becomes
\begin{align}
h_{ij}^{\mathrm{TT}}(t,\mathbf r) ={}& 4G \int\mathrm{d}^{3}y \int_{-\infty}^{+\infty}\mathrm{d}t'\, \frac{\delta(t-t'-R)}{R}\, T_{ij}^{\mathrm{TT}}(t',\mathbf y) \nonumber\\ &+ 8G\alpha \int\mathrm{d}^{3}y \int_{-\infty}^{+\infty}\mathrm{d}t'\, \frac{\delta''(t-t'-R)}{R}\, T_{ij}^{\mathrm{TT}}(t',\mathbf y) \nonumber\\ &- 2G\alpha \int\mathrm{d}^{3}y \int_{-\infty}^{+\infty}\mathrm{d}t'\, \delta'''(t-t'-R)\, T_{ij}^{\mathrm{TT}}(t',\mathbf y) + \mathcal{O}(\alpha^{2}).
\label{hij_convolution_explicit}
\end{align}
This form separates the general relativistic contribution from the two higher--derivative terms generated by the Ho\v{r}ava correction.

Let $d$ denote the characteristic size of the source and introduce the observation direction $\widehat{\mathbf n}=\mathbf r/r$. In the radiation zone, where $r\gg d$, the source--observer distance admits the expansion
\begin{equation}
R = r-\widehat{\mathbf n}\cdot\mathbf y + \mathcal{O}\!\left(\frac{d^{2}}{r}\right), \qquad \frac{1}{R} = \frac{1}{r} + \mathcal{O}\!\left(\frac{d}{r^{2}}\right).
\label{R_far_zone_expansion}
\end{equation}
In this manner,
\begin{equation}
t-t'-R = t_r-t' + \widehat{\mathbf n}\cdot\mathbf y + \mathcal{O}\!\left(\frac{d^{2}}{r}\right), \qquad t_r\equiv t-r.
\label{retarded_argument_expansion}
\end{equation}
For a slowly moving source, the characteristic wavelength is much larger than its size, so that $\varpi d\sim v\ll1$. Terms arising from $\widehat{\mathbf n}\cdot\mathbf y$ therefore belong to higher multipoles and finite--wavelength corrections. At leading quadrupolar order, we consistently set $R\simeq r$, and $t-t'-R\simeq t_r-t'$.

It is convenient to introduce the spatially integrated TT source,
\begin{equation}
\mathcal{T}_{ij}^{\mathrm{TT}}(t') \equiv \int\mathrm{d}^{3}y\, T_{ij}^{\mathrm{TT}}(t',\mathbf y).
\label{integrated_TT_source}
\end{equation}
At leading order in the radiation--zone and long--wavelength expansions, Eq.~\eqref{hij_convolution_explicit} reduces to
\begin{align}
h_{ij}^{\mathrm{TT}}(t,\mathbf r) \simeq{}& \frac{4G}{r} \int_{-\infty}^{+\infty}\mathrm{d}t'\, \delta(t_r-t')\, \mathcal{T}_{ij}^{\mathrm{TT}}(t') \nonumber\\ &+ \frac{8G\alpha}{r} \int_{-\infty}^{+\infty}\mathrm{d}t'\, \delta''(t_r-t')\, \mathcal{T}_{ij}^{\mathrm{TT}}(t') - 2G\alpha \int_{-\infty}^{+\infty}\mathrm{d}t'\, \delta'''(t_r-t')\, \mathcal{T}_{ij}^{\mathrm{TT}}(t') + \mathcal{O}(\alpha^{2}),
\label{hij_farzone_convolution}
\end{align}
up to higher multipoles and finite source corrections.

The required convolution can be evaluated through the general distributional identity
\begin{equation}
\int_{-\infty}^{+\infty}\mathrm{d}t'\, \delta^{(n)}(t_r-t')\,F(t') = \left. \frac{\mathrm{d}^{n}F(t')}{\mathrm{d}t'^{\,n}} \right|_{t'=t_r}. \label{delta_derivative_convolution} \end{equation} Indeed, setting $u=t_r-t'$ gives \begin{align} \int_{-\infty}^{+\infty}\mathrm{d}t'\, \delta^{(n)}(t_r-t')F(t') &= \int_{-\infty}^{+\infty}\mathrm{d}u\, \delta^{(n)}(u)F(t_r-u) \nonumber\\ &= (-1)^{n} \left. \frac{\mathrm{d}^{n}}{\mathrm{d}u^{n}} F(t_r-u) \right|_{u=0} = F^{(n)}(t_r).
\label{delta_derivative_proof}
\end{align}
The two signs generated by the distributional derivative and by differentiating $F(t_r-u)$ cancel. Applying Eq.~\eqref{delta_derivative_convolution} separately to the three terms in Eq.~\eqref{hij_farzone_convolution}, we obtain
\begin{align}
h_{ij}^{(0)\,\mathrm{TT}}(t,\mathbf r) &= \frac{4G}{r}\, \mathcal{T}_{ij}^{\mathrm{TT}}(t_r), \label{hij_GR_convolution_piece}\\ h_{ij}^{(2)\,\mathrm{TT}}(t,\mathbf r) &= \frac{8G\alpha}{r}\, \ddot{\mathcal{T}}_{ij}^{\mathrm{TT}}(t_r), \label{hij_delta2_piece}\\ h_{ij}^{(3)\,\mathrm{TT}}(t,\mathbf r) &= -2G\alpha\, \dddot{\mathcal{T}}_{ij}^{\mathrm{TT}}(t_r).
\label{hij_delta3_piece}
\end{align}
We have, therefore,
\begin{equation}
h_{ij}^{\mathrm{TT}}(t,\mathbf r) \simeq \frac{4G}{r}\, \mathcal{T}_{ij}^{\mathrm{TT}}(t_r) + \frac{8G\alpha}{r}\, \ddot{\mathcal{T}}_{ij}^{\mathrm{TT}}(t_r) - 2G\alpha\, \dddot{\mathcal{T}}_{ij}^{\mathrm{TT}}(t_r) + \mathcal{O}(\alpha^{2}).
\label{hij_integrated_source}
\end{equation}

The relation between the integrated spatial stress and the mass quadrupole follows directly from stress--energy conservation. Assuming a localized source for which surface terms vanish, we have $\partial_{\mu}T^{\mu\nu}=0$. Taking one time derivative of the quadrupole moment and using $\partial_tT^{00}=-\partial_kT^{k0}$ gives
\begin{align}
\dot I_{ij}(t) &= -\int\mathrm{d}^{3}y\, \partial_kT^{k0}(t,\mathbf y)\,y_i y_j = \int\mathrm{d}^{3}y\, \left[ T^{i0}(t,\mathbf y)y_j + T^{j0}(t,\mathbf y)y_i \right].
\label{quadrupole_first_derivative}
\end{align}
Differentiating once more, using $T^{0i}=T^{i0}$ and $\partial_tT^{0i}=-\partial_kT^{ki}$, and integrating by parts, we find
\begin{align}
\ddot I_{ij}(t) &= -\int\mathrm{d}^{3}y\, \left[ \partial_kT^{ki}(t,\mathbf y)y_j + \partial_kT^{kj}(t,\mathbf y)y_i \right] =  \int\mathrm{d}^{3}y\, \left[ T^{ji}(t,\mathbf y) + T^{ij}(t,\mathbf y) \right] = 2\int\mathrm{d}^{3}y\, T^{ij}(t,\mathbf y).
\label{quadrupole_second_derivative}
\end{align}
Since the TT projector is fixed by the observation direction in the radiation zone, it commutes with the spatial integral and the time derivatives. Thereby,
\begin{equation}
\mathcal{T}_{ij}^{\mathrm{TT}}(t) = \int\mathrm{d}^{3}y\, T_{ij}^{\mathrm{TT}}(t,\mathbf y) = \frac{1}{2}\, \ddot I_{ij}^{\mathrm{TT}}(t).
\label{quadrupole_relation_Tij}
\end{equation}
It follows immediately that
\begin{equation}
\ddot{\mathcal{T}}_{ij}^{\mathrm{TT}}(t) = \frac{1}{2}\, I_{ij}^{(4)\,\mathrm{TT}}(t), \qquad \dddot{\mathcal{T}}_{ij}^{\mathrm{TT}}(t) = \frac{1}{2}\, I_{ij}^{(5)\,\mathrm{TT}}(t).
\label{higher_quadrupole_relations}
\end{equation}
Substituting Eqs.~\eqref{quadrupole_relation_Tij} and \eqref{higher_quadrupole_relations} into Eq.~\eqref{hij_integrated_source} yields
\begin{equation}
h_{ij}^{\mathrm{TT}}(t,\mathbf r) \simeq \frac{2G}{r}\, I_{ij}^{(2)\,\mathrm{TT}}(t_r) + \frac{4G\alpha}{r}\, I_{ij}^{(4)\,\mathrm{TT}}(t_r) - G\alpha\, I_{ij}^{(5)\,\mathrm{TT}}(t_r) + \mathcal{O}(\alpha^{2}),
\label{strain_HL_expanded}
\end{equation}
or, equivalently,
\begin{equation}
h_{ij}^{\mathrm{TT}}(t,\mathbf r) \simeq \frac{2G}{r} \left[ I_{ij}^{(2)\,\mathrm{TT}} + 2\alpha I_{ij}^{(4)\,\mathrm{TT}} - \frac{\alpha r}{2}\, I_{ij}^{(5)\,\mathrm{TT}} \right]_{t=t_r} + \mathcal{O}(\alpha^{2}),
\label{strain_HL_checked}
\end{equation}
up to higher--multipole and finite--wavelength contributions. The standard general--relativistic quadrupole waveform is recovered when $\alpha\rightarrow 0$.

Projecting Eq.~\eqref{strain_HL_checked} onto the two polarization tensors introduced previously gives the observable strains
\begin{equation}
h_{A}(t,\mathbf r) \simeq \frac{2G}{r} \left[ I_{A}^{(2)} + 2\alpha I_{A}^{(4)} - \frac{\alpha r}{2}\, I_{A}^{(5)} \right]_{t=t_r}, \qquad A=+,\times,
\label{polarization_strains_HL}
\end{equation}
where the projection acts identically on every derivative because the isotropic $k^{4}$ operator neither mixes nor distinguishes the two tensor polarizations.

At fixed order in $\alpha$, Eqs.~\eqref{strain_HL_checked} and \eqref{polarization_strains_HL} remain local in the general relativistic retarded time. This property follows from the derivatives of the light cone delta distribution in Eq.~(\ref{greenfunddasdsa}); it is worth mentioning that it does not imply that the exact Ho\v{r}ava--Lifshitz propagator is nondispersive or strictly confined to $t=r$.

In particular, the fifth--derivative term should not be interpreted as an independent distance invariant radiative component. It represents the first--order expansion of the dispersive phase accumulated during propagation. For a Fourier component proportional to
$e^{-i\varpi t_r}$, Eq.~\eqref{strain_HL_checked} becomes
\begin{equation}
\widetilde h_{ij}^{\mathrm{TT}}(\varpi,r) \simeq \widetilde h_{ij}^{\mathrm{GR,TT}}(\varpi,r) \left( 1 - 2\alpha\varpi^{2} - \frac{i}{2}\alpha\varpi^{3}r \right).
\label{strain_HL_frequency_expanded}
\end{equation}
To the same perturbative order, this expression may be reorganized as
\begin{equation}
\widetilde h_{ij}^{\mathrm{TT}}(\varpi,r) \simeq \widetilde h_{ij}^{\mathrm{GR,TT}}(\varpi,r) \left(1-2\alpha\varpi^{2}\right) \exp\left( -\frac{i}{2}\alpha\varpi^{3}r \right).
\label{strain_HL_frequency_phase}
\end{equation}
The first factor produces a frequency dependent amplitude renormalization, whereas the exponential yields the accumulated phase correction
\begin{equation}
\delta\Psi_{\alpha}(\varpi,r) = -\frac{1}{2}\alpha\varpi^{3}r.
\label{HL_phase_shift}
\end{equation}
This result agrees with the perturbative wavenumber associated with the low--energy branch of the modified dispersion relation,
\begin{equation}
k(\varpi) = \varpi - \frac{1}{2}\alpha\varpi^{3} + \mathcal{O}(\alpha^{2}).
\label{HL_wavenumber_expansion}
\end{equation}
Accordingly, the apparently nondecaying term in Eq.~\eqref{strain_HL_expanded} is an important contribution generated by expanding the propagation phase. When this phase is retained in the exponential form of Eq.~\eqref{strain_HL_frequency_phase}, the waveform preserves its overall radiation--zone decay proportional to $1/r$. The local derivative expansion remains reliable provided that $|\alpha|\varpi^{2}\ll1$, so that  $|\alpha|\varpi^{3}r\ll1$. The first condition controls the effective field theory expansion of the dispersion relation, while the second controls the expansion of the accumulated propagation phase.


\subsection{Application with a binary black hole system}

\begin{figure}
    \centering
    \includegraphics[scale=0.5]{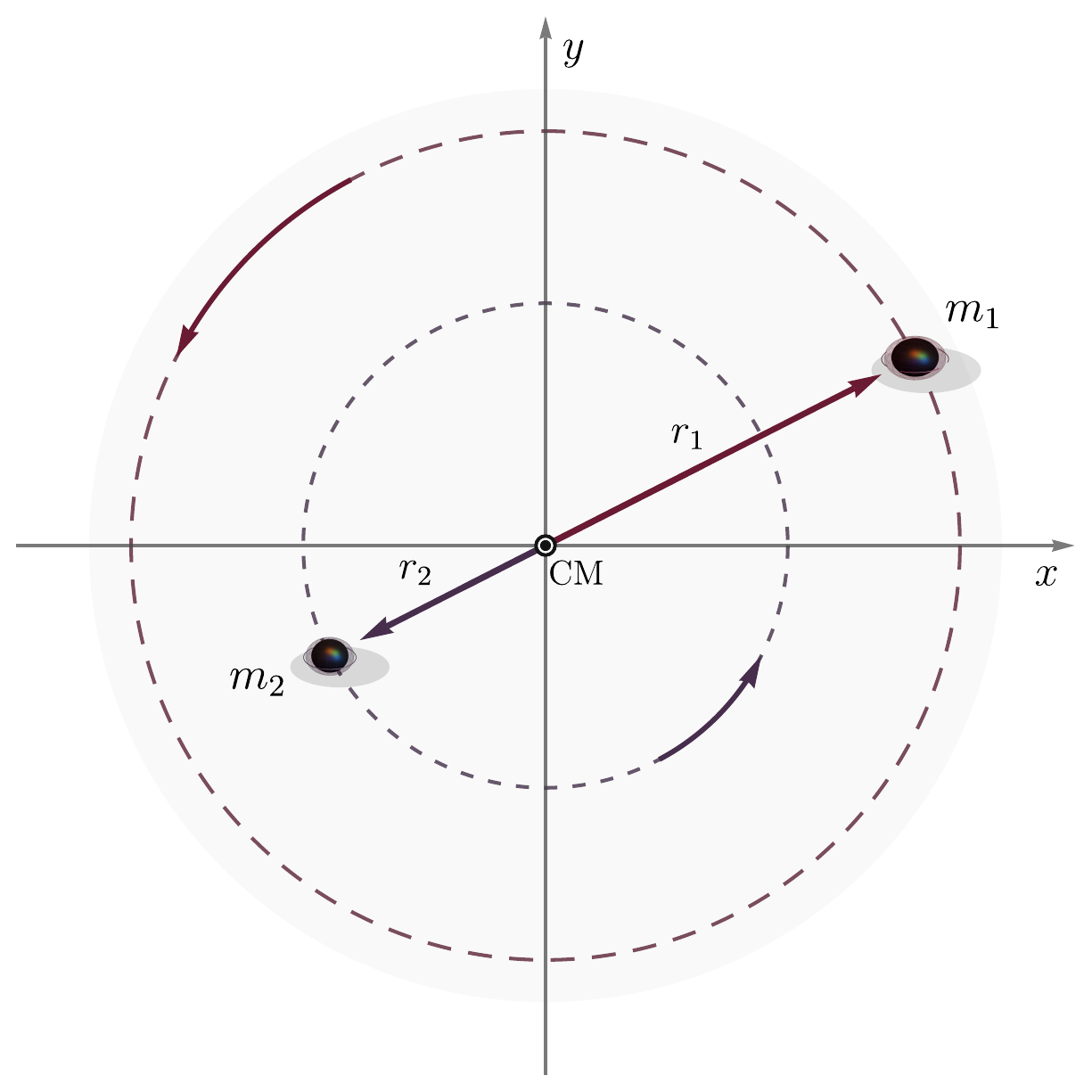}
    \caption{Illustration of a binary black-hole system in the barycentric
    frame. The compact objects, with masses $m_{1}$ and $m_{2}$, move on
    circular trajectories of radii $r_{1}$ and $r_{2}$ in the $xy$ plane.}
    \label{binary}
\end{figure}

To illustrate the Ho\v{r}ava--Lifshitz propagation corrections encoded in Eq.~\eqref{strain_HL_checked}, we apply the waveform formula to the binary black hole configuration shown in Fig.~\ref{binary}. The system consists of two compact objects with masses $m_{1}$ and $m_{2}$ moving on circular orbits in the $xy$ plane about their common barycenter. In the center of mass (CM) frame, the orbital radii are $r_{1}$ and $r_{2}$, and the binary separation is $l_{0}=r_{1}+r_{2}$. The total and reduced masses are
\begin{equation}
M=m_{1}+m_{2}, \qquad \mu=\frac{m_{1}m_{2}}{M},
\end{equation}
respectively.

The matter distribution is approximated by two pointlike sources confined to the orbital plane,
\begin{equation}
T_{00} = \delta(z) \Big[ m_{1}\delta(x-x_{1})\delta(y-y_{1}) + m_{2}\delta(x-x_{2})\delta(y-y_{2}) \Big].
\end{equation}
For uniform circular motion with orbital angular frequency $\omega$, their trajectories are
\begin{align}\label{eqmovbin}
x_{1}(t) &= \frac{m_{2}l_{0}}{M}\cos(\omega t), & y_{1}(t) &= \frac{m_{2}l_{0}}{M}\sin(\omega t), \nonumber\\ x_{2}(t) &= -\frac{m_{1}l_{0}}{M}\cos(\omega t), & y_{2}(t) &= -\frac{m_{1}l_{0}}{M}\sin(\omega t).
\end{align}

The corresponding mass quadrupole components are
\begin{align}\label{compI}
I_{xx}(t) &= \frac{\mu l_{0}^{2}}{2} \big[1+\cos(2\omega t)\big], \nonumber\\ I_{yy}(t) &= \frac{\mu l_{0}^{2}}{2} \big[1-\cos(2\omega t)\big], \nonumber\\ I_{xy}(t) = I_{yx}(t) &= \frac{\mu l_{0}^{2}}{2}\sin(2\omega t).
\end{align}
The trace free quadrupole, $Q_{ij}=I_{ij}-\frac{1}{3}\delta_{ij}I_{kk}$, differs from $I_{ij}$ only through time--independent terms. Since these terms vanish after differentiation, the radiative components may be written as
\begin{equation}
Q_{xx}(t) \simeq \frac{\mu l_{0}^{2}}{2}\cos(2\omega t), \qquad Q_{yy}(t) \simeq -\frac{\mu l_{0}^{2}}{2}\cos(2\omega t), \qquad Q_{xy}(t) = \frac{\mu l_{0}^{2}}{2}\sin(2\omega t).
\end{equation}

Introducing the phase $\Phi=2\omega t_{r}$, the derivatives required by Eq.~\eqref{strain_HL_checked} are
\begin{align}
Q_{xx}^{(2)} &= -2\mu l_{0}^{2}\omega^{2}\cos\Phi, & Q_{xx}^{(4)} &= 8\mu l_{0}^{2}\omega^{4}\cos\Phi, & Q_{xx}^{(5)} &= -16\mu l_{0}^{2}\omega^{5}\sin\Phi, \nonumber\\ Q_{yy}^{(n)} &= -Q_{xx}^{(n)}, && n=2,4,5, \nonumber\\ Q_{xy}^{(2)} &= -2\mu l_{0}^{2}\omega^{2}\sin\Phi, & Q_{xy}^{(4)} &= 8\mu l_{0}^{2}\omega^{4}\sin\Phi, & Q_{xy}^{(5)} &= 16\mu l_{0}^{2}\omega^{5}\cos\Phi.
\end{align}

For an observer located along the orbital normal, the displayed components coincide with their transverse--traceless projections. Substitution into Eq.~\eqref{strain_HL_checked} gives
\begin{align}\label{hxxnew}
h_{xx}^{\rm TT}(t,r) ={}& -\frac{4G\mu l_{0}^{2}\omega^{2}}{r}\cos\Phi + \frac{32G\alpha\mu l_{0}^{2}\omega^{4}}{r}\cos\Phi + 16G\alpha\mu l_{0}^{2}\omega^{5}\sin\Phi + \mathcal{O}(\alpha^{2}), \\ h_{yy}^{\rm TT}(t,r) ={}& +\frac{4G\mu l_{0}^{2}\omega^{2}}{r}\cos\Phi - \frac{32G\alpha\mu l_{0}^{2}\omega^{4}}{r}\cos\Phi - 16G\alpha\mu l_{0}^{2}\omega^{5}\sin\Phi + \mathcal{O}(\alpha^{2}), \\ h_{xy}^{\rm TT}(t,r) = h_{yx}^{\rm TT}(t,r) ={}& -\frac{4G\mu l_{0}^{2}\omega^{2}}{r}\sin\Phi + \frac{32G\alpha\mu l_{0}^{2}\omega^{4}}{r}\sin\Phi - 16G\alpha\mu l_{0}^{2}\omega^{5}\cos\Phi + \mathcal{O}(\alpha^{2}).
\label{hxynew}
\end{align}

It is useful to introduce the general relativistic amplitude
\begin{equation}
\mathcal{H}_{\rm GR} = \frac{4G\mu l_{0}^{2}\omega^{2}}{r}.
\end{equation}
Equations~\eqref{hxxnew}--\eqref{hxynew} can then be expressed as
\begin{align}
h_{xx}^{\rm TT} &= -\mathcal{H}_{\rm GR} \Big[ \left(1-8\alpha\omega^{2}\right)\cos\Phi - 4\alpha r\omega^{3}\sin\Phi \Big], \nonumber\\ h_{yy}^{\rm TT} &= -h_{xx}^{\rm TT}, \nonumber\\ h_{xy}^{\rm TT} = h_{yx}^{\rm TT} &= -\mathcal{H}_{\rm GR} \Big[ \left(1-8\alpha\omega^{2}\right)\sin\Phi + 4\alpha r\omega^{3}\cos\Phi \Big].
\label{binary_local_HL}
\end{align}

For the monochromatic binary harmonic $\varpi=2\omega$, the frequency domain waveform contains the factor
\begin{equation}
\left(1-2\alpha\varpi^{2}\right) \exp\left(-\frac{i}{2}\alpha\varpi^{3}r\right) = \left(1-8\alpha\omega^{2}\right) \exp\left(-4i\alpha\omega^{3}r\right).
\end{equation}
Keeping this accumulated phase exponentiated, the real waveform may be written in the phase resummed form
\begin{align}
h_{xx}^{\rm TT}(t,r) &\simeq -\mathcal{H}_{\rm GR} \left(1-8\alpha\omega^{2}\right) \cos\!\left(\Phi+\delta_{\alpha}\right), \nonumber\\ h_{yy}^{\rm TT}(t,r) &\simeq +\mathcal{H}_{\rm GR} \left(1-8\alpha\omega^{2}\right) \cos\!\left(\Phi+\delta_{\alpha}\right), \nonumber\\ h_{xy}^{\rm TT}(t,r) = h_{yx}^{\rm TT}(t,r) &\simeq -\mathcal{H}_{\rm GR} \left(1-8\alpha\omega^{2}\right) \sin\!\left(\Phi+\delta_{\alpha}\right),
\label{binary_resummed_HL}
\end{align}
where $\delta_{\alpha} = 4\alpha\omega^{3}r$. With the Fourier convention adopted previously, the corresponding frequency-domain phase correction is $\delta\Psi_{\alpha}=-4\alpha\omega^{3}r$.

Equation~\eqref{binary_resummed_HL} makes the physical content of the Ho\v{r}ava--Lifshitz correction transparent. The fourth--derivative term produces the frequency dependent amplitude factor $1-8\alpha\omega^{2}$, whereas the fifth--derivative term describes the dispersive phase accumulated during propagation. For positive $\alpha$, the amplitude is reduced within the perturbative regime $8|\alpha|\omega^{2}\ll1$, while negative $\alpha$ produces an enhancement. The phase correction grows linearly with the propagation distance and cubically with the orbital frequency.

The standard quadrupolar tensor structure is preserved. In particular, $h_{yy}^{\rm TT}=-h_{xx}^{\rm TT}$, while $h_{xy}^{\rm TT}$ remains in quadrature with the diagonal components. Both polarizations acquire the same amplitude renormalization and the same propagation phase, so there is no polarization mixing, helicity splitting, or additional oscillatory mode. The signal continues to oscillate at the binary harmonic $2\omega$. Nevertheless, its zero crossings are shifted relative to the general relativistic waveform because of the accumulated phase $\delta_{\alpha}$.

The parametric trajectory in the $\big(h_{xx}^{\rm TT},h_{yx}^{\rm TT}\big)$ plane satisfies
\begin{equation}
\left(h_{xx}^{\rm TT}\right)^{2} + \left(h_{yx}^{\rm TT}\right)^{2} \simeq \mathcal{H}_{\rm GR}^{2} \left(1-8\alpha\omega^{2}\right)^{2}.
\end{equation}
It therefore remains a circle centered at the origin. The Ho\v{r}ava--Lifshitz correction changes its radius but does not deform it into an ellipse. The common phase correction only changes the point occupied on the circle at a fixed retarded time and is consequently invisible when an unmarked complete orbit is displayed.

The deformation of a transverse ring of freely falling particles follows from the linearized mapping
\begin{equation}
X^{i} \longrightarrow X^{i} + \frac{1}{2}h^{i}{}_{j}X^{j}.
\end{equation}
At a fixed retarded phase, the magnitude of the stretching and compression is rescaled by $1-8\alpha\omega^{2}$. In addition, the common phase $\delta_{\alpha}$ rotates the principal axes of the instantaneous elliptic deformation by
\begin{equation}
\Delta\vartheta = \frac{\delta_{\alpha}}{2} = 2\alpha\omega^{3}r
\end{equation}
relative to the general relativistic configuration evaluated at the same retarded time.

\begin{figure}
    \centering
    \includegraphics[scale=0.6]{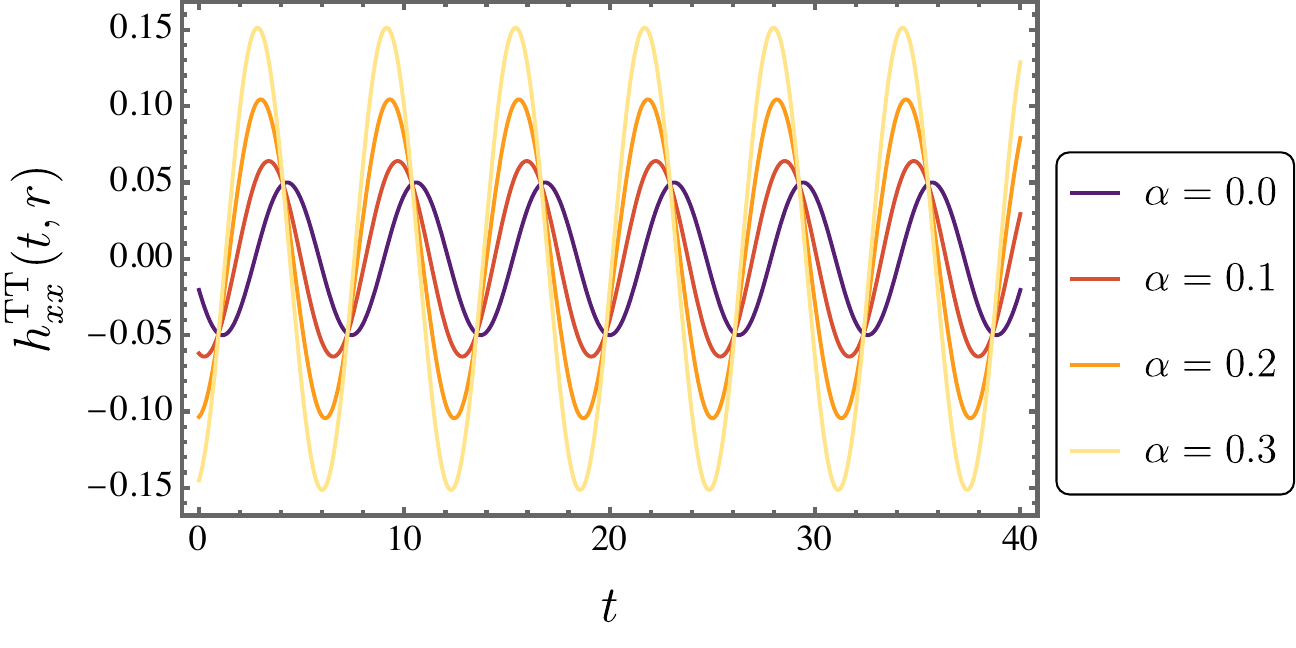}
    \caption{Time--domain waveform of the transverse--traceless component     $h_{xx}^{\rm TT}(t,r)$ for     $\alpha=0,\,0.1,\,0.2,\,0.3$, with     $G=1$, $\mu=1$, $l_{0}=1$, $\omega=0.5$, and $r=20$. The phase--resummed Ho\v{r}ava--Lifshitz correction preserves the oscillation frequency $2\omega$, while positive $\alpha$ progressively reduces the waveform amplitude through the factor $1-8\alpha\omega^{2}$ and shifts its phase by $\delta_{\alpha}=4\alpha\omega^{3}r$. Consequently, the maxima, minima, and zero crossings are displaced relative to the general relativistic waveform, although the oscillation period remains unchanged and no additional mode is generated.}
    \label{hxx_waveform}
\end{figure}

As illustrated in Fig.~\ref{hxx_waveform}, the Ho\v{r}ava--Lifshitz correction preserves the oscillation period of the quadrupolar signal. Within the directly expanded first-order waveform, increasing $\alpha$ enlarges the absolute excursions of $h_{xx}^{\rm TT}$ and displaces its maxima, minima, and zero crossings relative to the general relativistic curve. This apparent increase results from the combination of the cosine term with the additional sine contribution associated with the accumulated propagation phase. Since both contributions oscillate at the same binary harmonic $2\omega$, no additional frequency or oscillatory mode is introduced.

\begin{figure}
    \centering
    \includegraphics[scale=0.535]{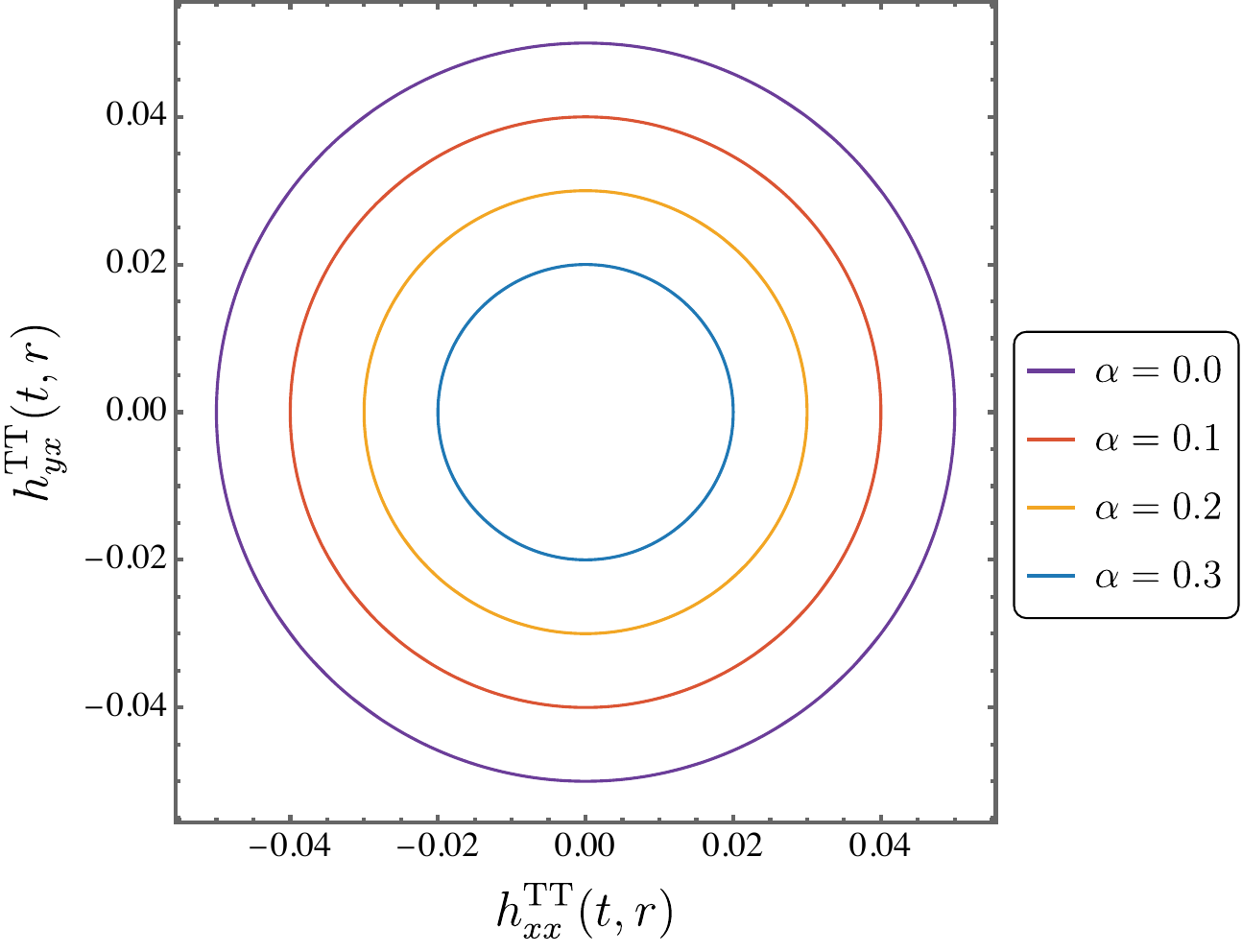}
    \caption{Parametric polarization trajectories in the $\big(h_{xx}^{\rm TT},h_{yx}^{\rm TT}\big)$ plane for     $\alpha=0,\,0.1,\,0.2,\,0.3$. The trajectories remain circles centered at the origin because the two transverse--traceless components retain equal amplitudes and a phase difference of $\pi/2$. Increasing $\alpha$ contracts their radius according to $H_{\rm GR}|1-8\alpha\omega^{2}|$, while the common propagation phase only changes the position occupied on each circle at a fixed retarded time. No polarization mixing or additional tensorial mode is introduced.}
    \label{paramtetricplotss}
\end{figure}

Figure~\ref{paramtetricplotss} displays the corresponding polarization orbits in the phase resummed representation. The trajectories remain circles centered at the origin, confirming that the two transverse--traceless components preserve equal amplitudes and remain in quadrature. Their radii decrease with increasing $\alpha$ according to the amplitude factor $|1-8\alpha\omega^{2}|$.

\begin{figure}
    \centering
    \includegraphics[scale=0.6]{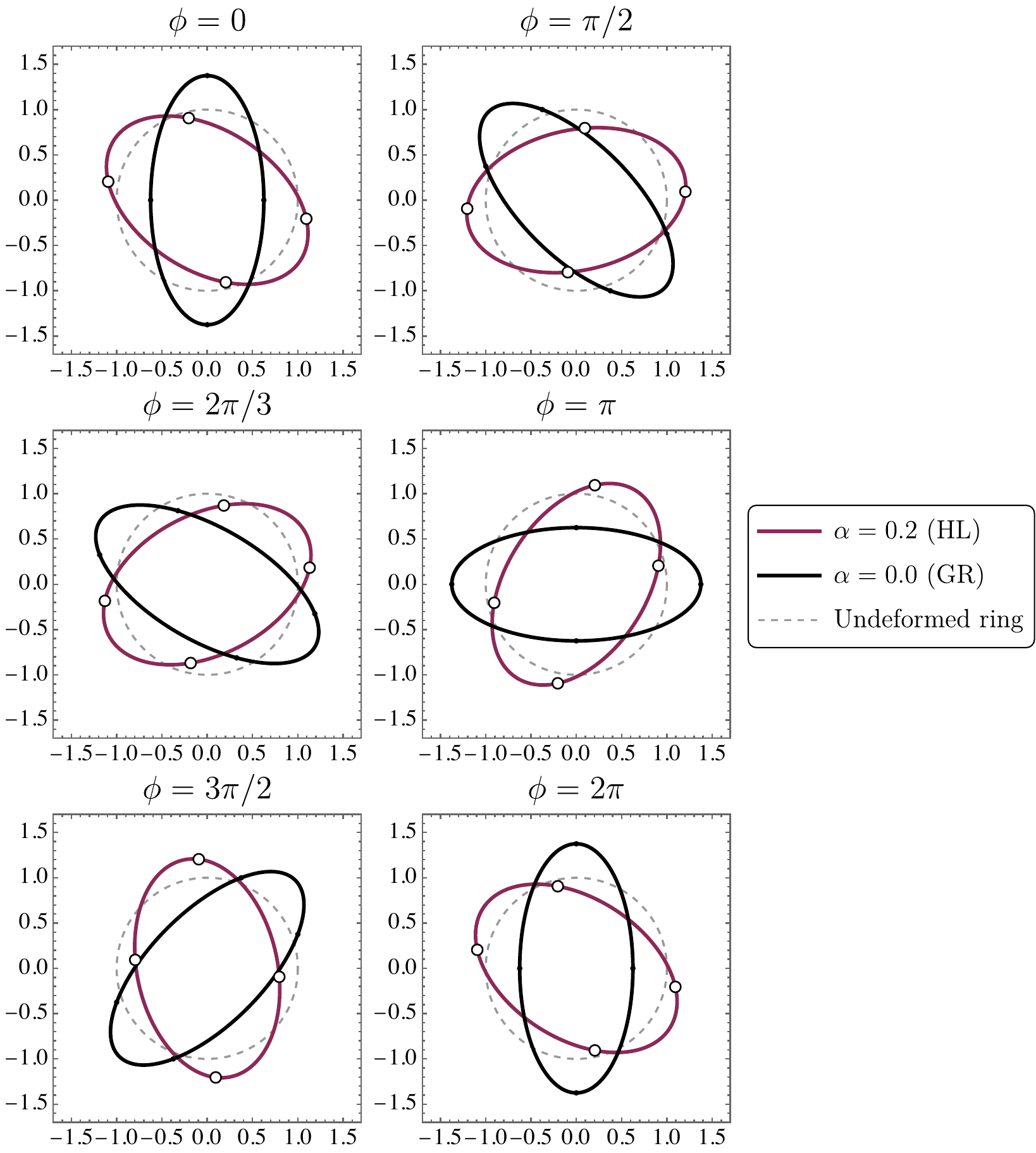}
    \caption{Transverse deformation of a ring of freely falling test particles induced by a gravitational wave in Ho\v{r}ava--Lifshitz     gravity. The six panels correspond to selected retarded phases $\phi=2\omega t_r$, with $t_r=t-r$, spanning one complete cycle. The dashed gray circle represents the undeformed configuration. The wine--colored curves show the Ho\v{r}ava--Lifshitz case with $\alpha=0.2$, whereas the black curves denote the general relativistic     reference with $\alpha=0$. White disks and black dots track particles initially located along the cardinal directions in the respective configurations. The parameters are $G=1$, $\mu=1$, $l_{0}=1$, $\omega=0.5$, and $r=20$. The Ho\v{r}ava--Lifshitz correction reduces the stretching and compression amplitude by the factor $1-8\alpha\omega^{2}$ and rotates the principal axes by $\Delta\vartheta=2\alpha\omega^{3}r$ relative to the general relativistic configuration at the same retarded phase.}
    \label{squeeze}
\end{figure}

The same phase resummed interpretation is visualized in Fig.~\ref{squeeze} through the transverse deformation of the particle ring. At a fixed retarded phase, the Ho\v{r}ava--Lifshitz configuration exhibits a smaller stretching and compression than the general relativistic reference, together with a rotation of the principal axes. After one complete oscillation, each configuration returns to its initial shape, consistently with the preservation of the harmonic frequency $2\omega$, as we mentioned before.

\section{Observable waveforms and inspiral evolution}

\subsection{Polarization waveforms for an arbitrary observation direction}

From Eq. \eqref{strain_HL_checked}, we can define $h_{ij}^{\mathrm{TT}}(t,\mathbf r)
\doteq
\frac{2G}{r}\,
\ddot I_{ij}^{\rm TT\, eff}(t_r)$ we define an effective $\ddot{I}_{ij}^{\rm TT\, eff}$, where
\begin{equation}\label{eq:eff_I}
    \ddot{I}^{\rm TT\, eff}_{ij}(t_r)=\ddot{I}_{ij}^{\rm TT}(t_r)
+
2\alpha
I_{ij}^{(4)\, \rm TT}(t_r)
-
\frac{\alpha}{2} r
I_{ij}^{(5)\, \rm TT}(t_r)\, ,
\end{equation}
where $I_{ij}(t)$ are given by \eqref{compI}.

Therefore, we can apply the techniques of the study of gravitational waves of general relativity to these effective quantities \cite{Maggiore:2007ulw}. By applying a rotation in Euler angles $(\theta,\phi)$ to a general  matrix $I_{ij}^{\rm eff}$, we can find the general form of the strain $h_{+}=h_{xx}^{\rm TT}=-h_{yy}^{\rm TT}$ and $h_{\times}=h_{xy}^{\rm TT}$ (see, for instance, \cite{Maggiore:2007ulw} for details)
\begin{align}
    h_{+}(t; \theta, \phi) = \frac{G}{r} &\left[\ddot{I}^{\rm TT\, eff}_{11}(\cos^{2} \phi - \sin^{2} \phi \cos^{2} \theta)+\ddot{I}^{\rm TT\, eff}_{22}(\sin^{2} \phi - \cos^{2} \phi \cos^{2} \theta)- \ddot{I}^{\rm TT\, eff}_{33} \sin^{2} \theta \right.\nonumber\\
    &\left.- \ddot{I}^{\rm TT\, eff}_{12} \sin 2\phi (1 + \cos^{2} \theta) 
    + \ddot{I}^{\rm TT\, eff}_{13} \sin \phi \sin 2\theta + \ddot{I}^{\rm TT\, eff}_{23} \cos \phi \sin 2\theta \right],
\end{align}

\begin{align}
    h_{\times}(t; \theta, \phi) = \frac{G}{r} &\left[ (\ddot{I}^{\rm TT\, eff}_{11} - \ddot{I}^{\rm TT\, eff}_{22}) \sin 2\phi \cos \theta + 2\ddot{I}^{\rm TT\, eff}_{12} \cos 2\phi \cos \theta - 2\ddot{I}^{\rm TT\, eff}_{13} \cos \phi \sin \theta \right.\nonumber\\
    &\left.+ 2\ddot{I}^{\rm TT\, eff}_{23} \sin \phi \sin \theta \right].
\end{align}

Since the only non-zero $\ddot{I}^{\rm eff}_{ij}$ are $\ddot{I}^{\rm TT\, eff}_{11}$, $\ddot{I}^{\rm TT\, eff}_{22}$ and $\ddot{I}^{\rm TT\, eff}_{12}$, and using Eqs.\eqref{eq:eff_I} and \eqref{compI}, we derive the following strains in a general direction $(\theta,\phi)$
\begin{align}
   h_{+}(t; \theta, \phi) &= -\frac{1}{r}4 G \mu  l_0^2 \omega ^2 \left(\frac{1+\cos ^2\theta}{2}\right)\left[\left(1-8\alpha \omega ^2\right) \cos (2 (\omega t_r +\phi ))-4 \alpha  r \omega ^3 \sin (2 (t_r \omega +\phi ))\right]\, ,\\
   h_{\times}(t; \theta, \phi) &=-\frac{1}{r}4 G \mu  l_0^2 \omega ^2 \cos \theta  \left[\left(1-8 \alpha  \omega ^2\right) \sin (2 (\omega t_r +\phi ))+4 \alpha  r \omega ^3 \cos (2 (t_r \omega +\phi ))\right]\, .
\end{align}

We can shift the origin of time such that $t_r\mapsto t_r+\pi/(2\omega)$, such that $\cos(2(\omega t_r+\phi))\mapsto -\cos(2(\omega t_r+\phi))$ and $\sin(2(\omega t_r+\phi))\mapsto -\sin(2(\omega t_r+\phi))$. We also identify the angle $\theta$ with the angle, that we call $\iota$, between the normal to the orbit and the line-of-sight that connects the binary system and our laboratory.

The Newtonian two-body dynamics can be reduced to the one-body problem with mass equal to the reduced mass $\mu$ and equation of motion $\ddot{\vec{r}}=-(GM/l_0^3)\vec{r}$, where $M$ is the total mass, and $\vec{r}=\vec{r}_2-\vec{r}_1$ where $|\vec{r}|=l_0$. The orbital frequency $\omega$ is related $M$ and $l_0$ by noticing that the linear speed of the center of mass frame $v=\omega l_0$ and from the centripetal force $v^2/l_0=GM/l_0^2$, which gives
\begin{equation}\label{eq:w-l0}
    \omega^2=\frac{GM}{l_0^3}\, .
\end{equation}

If we use the chirp mass $M_c=\mu^{3/5}M^{2/5}$, and the frequency of the gravitational wave $f_{\rm gw}=\omega_{\rm gw}/(2\pi)$, where $\omega_{\rm gw}=2\omega$ (twice the frequency of the orbit), we can express our result as

\begin{align}
    h_{+}(t; \theta, \phi) &=\frac{4}{r}\left(GM_c\right)^{5/3}\left(\pi f_{\rm gw}\right)^{2/3}\left(\frac{1+\cos ^2\iota}{2}\right)
    \begin{aligned}[t]
    &\Bigl[\left(1-8 \alpha  \omega ^2\right) \cos (2\pi f_{\rm gw} t_r +2\phi)\\
    &\qquad-4 \alpha  r \omega ^3\sin (2\pi f_{\rm gw} t_r +2\phi)\Bigr]\, ,
    \end{aligned}\label{strain_cross_fixed}\\
    h_{\times}(t; \theta, \phi) &=\frac{4}{r}\left(GM_c\right)^{5/3}\left(\pi f_{\rm gw}\right)^{2/3} \cos \iota
    \begin{aligned}[t]
    &\Bigl[\left(1-8 \alpha  \omega ^2\right) \sin (2\pi f_{\rm gw} t_r +2\phi)\\
    &\qquad+4 \alpha  r \omega ^3 \cos (2\pi f_{\rm gw} t_r +2\phi)\Bigr]\, .
    \end{aligned}\label{strain_x_fixed}
\end{align}


\subsection{Polarization waveforms for an arbitrary observation direction}

Having obtained the polarization waveforms for an arbitrary observation direction, we calculate the corresponding gravitational wave energy flux. The power emitted per unit solid angle is determined by the temporal average of the squared time derivatives of the two tensor polarizations.

The power radiated per unit solid angle is \cite{Maggiore:2007ulw}
\begin{align}
    \frac{\mathrm{d}P}{\mathrm{d}\Omega}&=\frac{r^2}{32\pi G}\left\langle \dot{h}_{ij}^{\rm TT}\dot{h}_{ij}^{\rm TT}\right\rangle=\frac{r^2}{16\pi G}\left\langle \left(\dot{h}_{+}\right)^2+\left(\dot{h}_{\times}\right)^2\right\rangle\, ,
\end{align}
where $\langle ,\rangle$ is a temporal average over several periods of the gravitational wave and $\dot{h}_{ij}^{\rm TT}$ is calculated at the retarded time $t_r$. We use the fact that $\langle \cos^2(2\pi f_{\rm gw} t_r +2\phi)\rangle=\langle \cos^2(2\pi f_{\rm gw} t_r +2\phi)\rangle=1/2$ and $\langle \cos(2\pi f_{\rm gw} t_r +2\phi)\sin(2\pi f_{\rm gw} t_r +2\phi)\rangle=0$ to find that
\begin{equation}
    \frac{\mathrm{d}P}{\mathrm{d}\Omega}=\frac{2}{\pi G}\left(GM_c\pi f_{\rm gw}\right)^{10/3}\left[\left(\frac{1+\cos^2\iota}{2}\right)^2+\cos^2\iota\right]\left(1-16\pi^2\alpha f_{\rm gw}^2\right)\, .
\end{equation}

Performing an integration over the solid angle, we derive the total radiated power
\begin{equation}
    P_{\rm total}=\frac{32}{5G}\left(GM_c f_{\rm gw}\right)^{10/3}\left(1-16\pi^2\alpha f_{\rm gw}^2\right)\, .
\end{equation}

We notice that the term proportional to $a_3$ does not contribute to this quantity since it averages out after a complete orbit in this approximation. It is possible that they contribute in the post--Newtonian regime, which goes beyond the scope of this paper.


\subsection{Adiabatic inspiral and chirp evolution}

The loss of orbital energy causes the binary separation to decrease and the gravitational wave frequency to increase gradually. We now determine the leading Ho\v{r}ava--Lifshitz correction to this adiabatic inspiral evolution by imposing energy balance between the orbital-energy loss and the total gravitational wave luminosity.

Due to the energy emission, the radius of the orbit must decrease in time. Using \eqref{eq:w-l0}, we see that $\dot{l}_0=-2\omega l_0\dot{\omega}/(3\omega^{2})$, which implies that as long as the condition $\dot{\omega}\ll \omega^2$, we have $\dot{l}_0\ll \omega l_0$ (the tangential velocity) and the approximation of circular orbit with slowly decreasing radius is allowed. From Newtonian mechanics \cite{Goldstein:2002}, the energy of the orbit is given by $E_{\rm orb}=-Gm_1m_2/(2l_0)=-(G^2M_c^5\omega_{\rm gw})^{1/3}$, where we used the definition of the chirp mass, the relation between $l_0$ and $\omega$ \eqref{eq:w-l0} and its relation with $\omega_{\rm gw}=2\omega$. Equating $P_{\rm total}=-\mathrm{d}E_{\rm orb}/\mathrm{d}t_r$, we find
\begin{equation}
\dot{f}_{\rm gw}=\frac{96}{5}\pi^{8/3}\left(G M_c\right)^{5/3}f_{\rm gw}^{11/3}(1-16 \pi ^2 \alpha  f_{\rm gw}^2)\, .
\end{equation}

The real solution of this equation is
\begin{equation}\label{fgw_t}
f_{\rm gw}(t)=\frac{1}{\pi\left(G M_c\right)^{5/8} }\left(\frac{5}{256 \tau}\right)^{3/8}\left[1+\frac{3\alpha}{8\left(GM_c\right)^{5/4}}\left(\frac{5}{\tau}\right)^{3/4}\right]\, ,
\end{equation}
where $\tau=t_{\rm coal}-t$ is the time remaining until coalescence, determined by the coalescence time $t_{\rm coal}$. Notice that $\tau$ reads the same whether we consider the retarded time and retarded coalescence time or the observer times. When $t=t_{\rm coal}$, the frequency and the amplitude of the waves diverge (from \eqref{strain_cross_fixed} and \eqref{strain_x_fixed}) which means that we have entered a regime in which neither the linear approximation, nor the low tangential velocity approximation are valid.

\begin{figure}[t]
    \centering
    \includegraphics[scale=0.6]{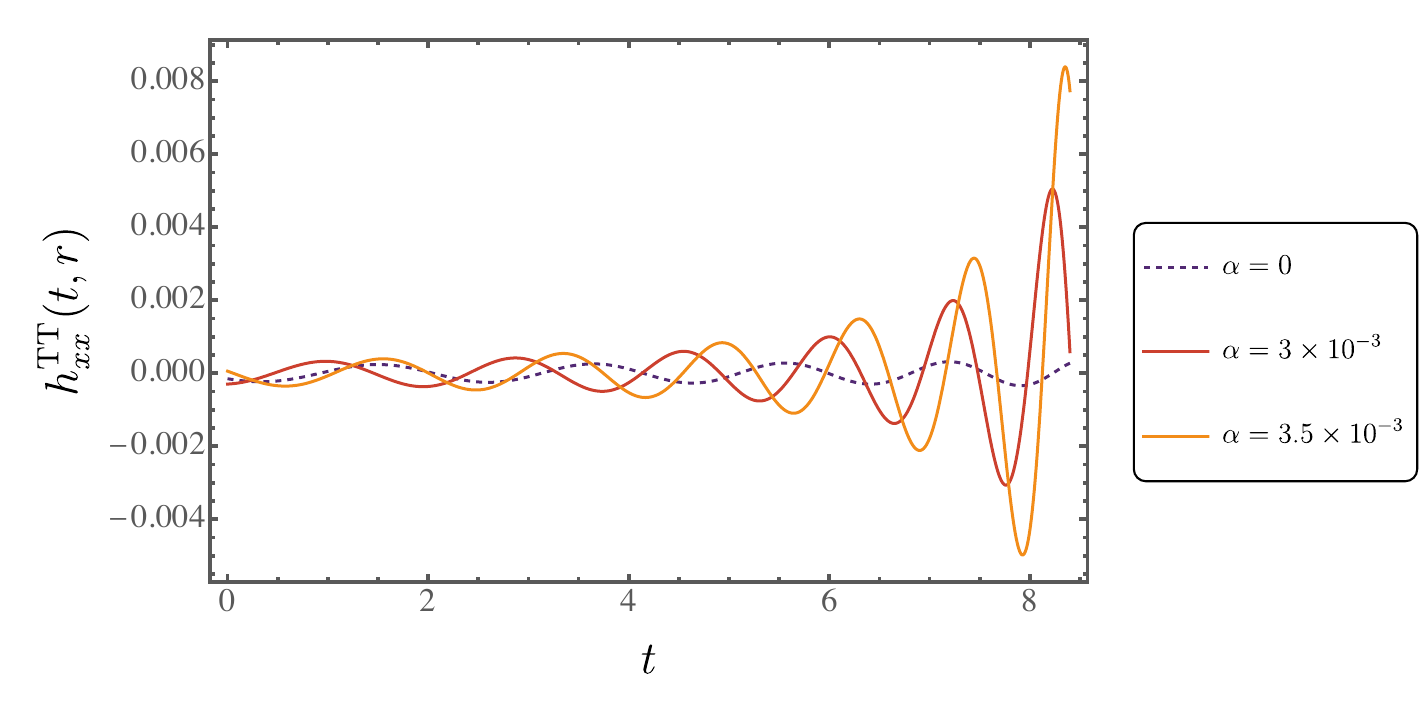}
    \caption{Time evolution of the gravitational wave strain $h_{xx}^{\rm TT}(t,r)$ for three values of the quantum-gravity parameter $\alpha$. The dashed purple curve corresponds to the standard general relativity prediction ($\alpha=0$), while the solid red and orange curves show the modified waveforms for $\alpha=3\times10^{-3}$ and $\alpha=3.5\times10^{-3}$, respectively. The parameters are fixed to $\iota=0$, $G=1$, $r=20$, $t_{\rm coal}=10$, and $M_c=0.0155$. The modifications become more pronounced as the binary approaches coalescence.}
    \label{fig:hxx_inspiral}
\end{figure}

To study this regime, we consider the effect of the change in the wave frequency, which can be found by replacing $f_{\rm gw}\mapsto f_{\rm gw}(t)$ from \eqref{fgw_t} and $(2\pi f_{\rm gw} t_{r}+2\phi)\mapsto \Phi(t_r)$ from \eqref{Phi_t} in \eqref{strain_cross_fixed} and \eqref{strain_x_fixed}, where
\begin{equation}\label{Phi_t}
\Phi(t)=\int^{t}_{t_{\rm coal}} 2\pi f_{\rm gw}(t')\mathrm{d}t'=-\frac{2}{\left(5 G M_c\right){}^{5/8}} \tau^{5/8}\left(1-\frac{15\alpha}{8(GM_c)^{5/4}}\left(\frac{5}{\tau}\right)^{3/4}\right)+\Phi_0\, ,
\end{equation}
and $\Phi_0=\Phi(t_{\rm coal})$. We depict this behavior in Fig.~\ref{fig:hxx_inspiral}, in which we observe the effect of the Lorentz violating parameter $\alpha$ in the dephasing and modification of the amplitude of the wave $h_{xx}^{\rm TT}=h_{+}$. We notice that the Lorentz violating effect becomes more pronounced as we approach the coalescence time, and it anticipates the drastic growth of the amplitudes when approaching the coalescence time, which can be inferred by \eqref{fgw_t}.

The time scale scale in which the corrections are of the order of the undeformed term satisfies $\tau\sim \alpha^{4/3}(GM_c)^{-5/3}$. If Planck scale effects are responsible for violating the Lorentz symmetry, which means that $\alpha \sim G$ (the square of the Planck length in natural units), than this effect would be relevant for $\tau\sim t_{\rm Pl}(M_{\rm Pl}/M_c)^{5/3}$, where $t_{\rm Pl}=\sqrt{G}$ is the Planck time and $M_{\rm Pl}=t_{\rm P}^{-1}$ is the Planck mass. However, in this regime, the linear approximation for gravitational waves is no longer valid and a more detailed analysis should be carried out.

\section{Observational constraints on the Ho\v{r}ava--Lifshitz
dispersion parameter}
\label{sec:bounds_LVK}

The most recent cumulative LIGO--Virgo--KAGRA transient catalog is
GWTC--5.0, which includes events detected during the second part of the
fourth observing run \cite{LVK:GWTC5}. At the time of writing, however,
a catalog-wide modified-dispersion analysis incorporating the GWTC--5.0
events has not yet been released. The most recent public constraints on
the modified gravitational-wave dispersion relation are therefore those
obtained from the GWTC--4.0 parameterized tests of general relativity
\cite{LVK:GWTC4TestsII}. We use these results below and do not
extrapolate them to the larger GWTC--5.0 sample.


\subsection{Mapping to the LVK modified-dispersion framework}
\label{subsec:LVK_mapping}

The tensor modes considered in this work obey \begin{equation} 
\omega^{2}=k^{2}+\alpha k^{4}, \label{eq:HL_dispersion_bounds} 
\end{equation} 
where $[\alpha]=L^{2}$ and $c=\hbar=1$. To avoid confusing the Ho\v{r}ava--Lifshitz coefficient $\alpha$ with the exponent used in the LVK parametrization, we denote the latter by $\beta$. The LVK modified dispersion relation is written as
\begin{equation}
E^{2} = k^{2}c^{2} + A_{\beta}\,k^{\beta}c^{\beta},
\label{eq:LVK_MDR_beta}
\end{equation}
where $A_{\beta}$ determines the magnitude of the modification \cite{Mirshekari:2011yq,Ezquiaga:2022MDR,Baka:2025MDR}. In natural units, comparison of Eqs.~\eqref{eq:HL_dispersion_bounds} and \eqref{eq:LVK_MDR_beta} gives
\begin{equation}
\beta=4, \qquad A_{4}=\alpha.
\label{eq:A4_alpha_mapping}
\end{equation}

For a source at redshift $z$, the LVK analysis introduces the frequency domain propagation phase
\begin{equation}
\delta\Psi_{\beta}^{\mathrm{prop}}(f) = - \frac{ \pi D_{\beta}\,h_{\mathrm{P}}^{\,\beta-2} }{ (1+z)^{\beta-1}c } A_{\beta}f^{\beta-1},
\label{eq:LVK_phase_general}
\end{equation}
where $h_{\mathrm{P}}$ is Planck's constant and
\begin{equation}
D_{\beta}(z) = \frac{ c(1+z)^{1-\beta} }{ H_{0} } \int_{0}^{z} \frac{ (1+z')^{\beta-2} }{ \sqrt{ \Omega_{m}(1+z')^{3}+\Omega_{\Lambda} } } \,\mathrm{d}z'
\label{eq:LVK_modified_distance}
\end{equation}
is the modified cosmological distance
\cite{Mirshekari:2011yq,Ezquiaga:2022MDR,LVK:GWTC4TestsII}.

For $\beta=4$, Eq.~\eqref{eq:LVK_phase_general} becomes
\begin{equation}
\delta\Psi_{4}^{\mathrm{prop}}(f) = - \frac{ \pi D_{4}\,h_{\mathrm{P}}^{2} }{ (1+z)^{3}c } \alpha f^{3}.
\label{eq:LVK_phase_beta4}
\end{equation}
In the local limit, $z\rightarrow0$ and $D_{4}\rightarrow r$. Setting $c=\hbar=1$, for which $h_{\mathrm{P}}=2\pi$, gives
\begin{equation}
\delta\Psi_{4}^{\mathrm{prop}}(f) \longrightarrow -4\pi^{3}\alpha f^{3}r = -\frac{\alpha}{2}(2\pi f)^{3}r.
\label{eq:local_phase_beta4}
\end{equation}
This is precisely the accumulated phase obtained from the phase resummed waveform derived in the previous sections. In particular, for either tensor polarization,
\begin{equation}
\widetilde{h}_{A}(f,r) \simeq \widetilde{h}^{\mathrm{GR}}_{A}(f,r) \left(1-8\pi^{2}\alpha f^{2}\right) \exp\left[-4i\pi^{3}\alpha f^{3}r\right], \qquad A=+,\times,
\label{eq:HL_detected_waveform_bounds}
\end{equation}
to first order in the local amplitude correction, while retaining the accumulated propagation phase in exponential form. Consequently, the term proportional to $\alpha r f^{3}$ is not an independent distance invariant radiative contribution. It is the first=-order expansion of the phase in Eq.~\eqref{eq:HL_detected_waveform_bounds}.


\subsection{GWTC--4.0 constraint}
\label{subsec:GWTC4_alpha_bound}

The GWTC--4.0 modified--dispersion analysis combines 83 events and finds no evidence for dispersive gravitational wave propagation \cite{LVK:GWTC4TestsII}. The reported dimensionless parameter is
\begin{equation}
\overline{A}_{\beta} \equiv \frac{A_{\beta}}{\mathrm{eV}^{\,2-\beta}}.
\label{eq:Abar_definition}
\end{equation}
For $\beta=4$, the combined 90\% credible interval is
\begin{equation}
\overline{A}_{4} \in \left[-0.62,\,0.19\right]\times10^{3}.
\label{eq:GWTC4_A4_interval}
\end{equation}
Using Eq.~\eqref{eq:A4_alpha_mapping}, the corresponding constraint on the Ho\v{r}ava--Lifshitz coefficient is
\begin{equation}
{ -6.2\times10^{2}\ \mathrm{eV}^{-2} < \alpha < 1.9\times10^{2}\ \mathrm{eV}^{-2} } \qquad (90\%~\text{credible interval}).
\label{eq:alpha_GWTC4_eV}
\end{equation}

For comparison, the reanalysis of the GWTC--3 events using the same group velocity prescription gave
\begin{equation}
\overline{A}_{4}^{\mathrm{GWTC-3}} \in \left[-1.69,\,0.64\right]\times10^{3}
\label{eq:GWTC3_reanalysis}
\end{equation}
at 90\% credibility \cite{Baka:2025MDR,LVK:GWTC4TestsII}.
The width of the $\beta=4$ interval is therefore reduced by approximately a factor of $2.9$ after including the O4a events. The older bounds obtained using the particle velocity prescription should not be combined directly with Eqs.~\eqref{eq:GWTC4_A4_interval} and \eqref{eq:GWTC3_reanalysis}, since the current LVK analysis employs the group velocity, which agrees with the WKB description of wave propagation \cite{Ezquiaga:2022MDR,Baka:2025MDR}. 
Using, $1~\mathrm{eV}^{-1} = 1.97327\times10^{-7}\ \mathrm{m} = 6.58212\times10^{-16}\ \mathrm{s}$,
Eq.~\eqref{eq:alpha_GWTC4_eV} becomes
\begin{equation}
-2.41\times10^{-11}\ \mathrm{m}^{2} < \alpha < 7.40\times10^{-12}\ \mathrm{m}^{2}
\label{eq:alpha_GWTC4_m2}
\end{equation}
or, equivalently,
\begin{equation}
-2.69\times10^{-28}\ \mathrm{s}^{2} < \alpha < 8.23\times10^{-29}\ \mathrm{s}^{2}.
\label{eq:alpha_GWTC4_s2}
\end{equation}
A sign independent envelope of the central 90\% interval is
\begin{equation}
|\alpha| \lesssim 6.2\times10^{2}\ \mathrm{eV}^{-2} \simeq 2.41\times10^{-11}\ \mathrm{m}^{2}.
\label{eq:absolute_alpha_envelope}
\end{equation}
Introducing the length and energy scales
\begin{equation}
\ell_{\alpha}\equiv\sqrt{|\alpha|}, \qquad \Lambda_{\alpha}\equiv\frac{1}{\sqrt{|\alpha|}},
\label{eq:alpha_scales}
\end{equation}
we have therefore
\begin{equation}
\ell_{\alpha} \lesssim 4.91\times10^{-6}\ \mathrm{m}, \qquad \Lambda_{\alpha} \gtrsim 4.02\times10^{-2}\ \mathrm{eV}.
\label{eq:alpha_length_energy_bounds}
\end{equation}

For comparison with calculations in which the compact object mass is used to normalize the coupling, we may define
\begin{equation}
\widehat{\alpha}_{M} \equiv \frac{\alpha}{(GM/c^{2})^{2}}.
\label{eq:dimensionless_alpha_mass}
\end{equation}
The conservative GWTC--4.0 envelope then gives
\begin{equation}
|\widehat{\alpha}_{M}| \lesssim 1.11\times10^{-19} \left( \frac{10M_{\odot}}{M} \right)^{2}.
\label{eq:dimensionless_alpha_bound}
\end{equation}


\subsection{Validity of the perturbative expansion}
\label{subsec:alpha_validity_bound}

The local derivative expansion requires
\begin{equation}
|\alpha|(2\pi f)^{2}\ll1.
\label{eq:local_EFT_condition_bounds}
\end{equation}
Using the conservative value in Eq.~\eqref{eq:alpha_GWTC4_s2}, at $f=10^{3}\,\mathrm{Hz}$ we are able write
\begin{equation}
|\alpha|(2\pi f)^{2} \lesssim 1.1\times10^{-20}.
\label{eq:EFT_condition_numerical}
\end{equation}
The derivative expansion of the dispersion relation is, in other words, exceptionally well controlled throughout the LVK band.

This conclusion does not imply that the accumulated propagation phase must also be expanded. The relevant phase parameter contains the source distance,
\begin{equation}
\left| \delta\Psi_{4}^{\mathrm{prop}} \right| \sim 4\pi^{3}|\alpha|f^{3}r,
\label{eq:accumulated_phase_parameter}
\end{equation}
and can remain observable over astrophysical baselines even when Eq.~\eqref{eq:local_EFT_condition_bounds} is satisfied by many orders of magnitude. The exponential representation in Eq.~\eqref{eq:HL_detected_waveform_bounds}, or its cosmological generalization in Eq.~\eqref{eq:LVK_phase_beta4}, must consequently be used in parameter estimation.

For $\alpha<0$, the truncated dispersion relation also requires
\begin{equation}
1+\alpha k^{2}>0, \qquad\text{or equivalently}\qquad k<|\alpha|^{-1/2}.
\label{eq:negative_alpha_stability}
\end{equation}
The GWTC--4.0 interval satisfies this condition throughout the detector band. Nevertheless, a negative $k^{4}$ coefficient cannot define the ultraviolet completion by itself; stability at higher momenta must be supplied by the omitted operators, including the $k^{6}$ contribution of the complete Ho\v{r}ava--Lifshitz dispersion relation \cite{Horava:2009uw}.


\subsection{Multimessenger speed constraint}
\label{subsec:GW170817_alpha}

The group velocity associated with Eq.~\eqref{eq:HL_dispersion_bounds} is
\begin{equation}
v_{g} = \frac{\mathrm{d}\omega}{\mathrm{d}p} = \frac{1+2\alpha k^{2}}{\sqrt{1+\alpha k^{2}}} \simeq 1+\frac{3}{2}\alpha k^{2},
\label{eq:group_velocity_alpha_bounds}
\end{equation}
where the last expression applies when $|\alpha|k^{2}\ll1$. Expressing $\alpha$ in units of time squared and setting $k\simeq2\pi f$, we have
\begin{equation}
\frac{v_{g}-c}{c} \simeq \frac{3}{2}\alpha(2\pi f)^{2}.
\label{eq:speed_alpha_relation}
\end{equation}

The association between GW170817 and GRB\,170817A constrains the relative propagation speed to
\begin{equation}
-3\times10^{-15} \lesssim \frac{v_{g}-c}{c} \lesssim 7\times10^{-16}
\label{eq:GW170817_speed_interval}
\end{equation}
after accounting conservatively for the possible after accounting conservatively for the possible delay between the binary merger and gamma ray emission \cite{LIGOScientific:2017zic}. For a representative effective frequency $f_{\mathrm{eff}}$, Eqs.~\eqref{eq:speed_alpha_relation} and \eqref{eq:GW170817_speed_interval} imply
\begin{equation}
-5.1\times10^{-21} \left( \frac{100\,\mathrm{Hz}}{f_{\mathrm{eff}}} \right)^{2} \mathrm{s}^{2} \lesssim \alpha \lesssim 1.2\times10^{-21} \left( \frac{100\,\mathrm{Hz}}{f_{\mathrm{eff}}} \right)^{2} \mathrm{s}^{2}.
\label{eq:alpha_multimessenger_s2}
\end{equation}
At $f_{\mathrm{eff}}=100\,\mathrm{Hz}$, this corresponds to
\begin{equation}
-4.6\times10^{-4}\ \mathrm{m}^{2} \lesssim \alpha \lesssim 1.1\times10^{-4}\ \mathrm{m}^{2}.
\label{eq:alpha_multimessenger_m2}
\end{equation}
This result is several orders of magnitude weaker than the GWTC--4.0 accumulated phase constraint. It nevertheless provides an independent multimessenger consistency test that does not rely on combining a population of compact binary signals.


\subsection{Constraints implied by the waveform and inspiral dynamics}
\label{subsec:source_bounds_alpha}

The waveform derived in the previous sections contains local modifications of the amplitude, luminosity, and chirp evolution in addition to the propagation phase. Written in terms of the observed gravitational wave frequency, their leading fractional corrections are
\begin{align}
\frac{\Delta h_{A}}{h_{A}^{\mathrm{GR}}} &= -8\pi^{2}\alpha f^{2}, \label{eq:amplitude_fractional_alpha} \\ \frac{\Delta P}{P_{\mathrm{GR}}} &= -16\pi^{2}\alpha f^{2},
\label{eq:power_fractional_alpha}
\\
\frac{\Delta\dot{f}}{\dot{f}_{\mathrm{GR}}} &= -16\pi^{2}\alpha f^{2}.
\label{eq:chirp_fractional_alpha}
\end{align}
In particular, the corrected total luminosity and chirp rate are
\begin{align}
P &= \frac{32}{5G} \left( \pi G\mathcal{M}_{c}f \right)^{10/3} \left( 1-16\pi^{2}\alpha f^{2} \right), \label{eq:corrected_power_bounds} \\ \dot{f} &= \frac{96}{5} \pi^{8/3} (G\mathcal{M}_{c})^{5/3} f^{11/3} \left( 1-16\pi^{2}\alpha f^{2} \right).
\label{eq:corrected_chirp_bounds}
\end{align}

If observations restrict the fractional amplitude, luminosity, or chirp rate deviations at a characteristic frequency $f_{\ast}$ to $\varepsilon_{h}$, $\varepsilon_{P}$, and $\varepsilon_{\dot f}$, respectively, the corresponding translations are
\begin{align}
|\alpha| &\lesssim \frac{\varepsilon_{h}} {8\pi^{2}f_{\ast}^{2}}, \label{eq:alpha_amplitude_generic} \\ |\alpha| &\lesssim \frac{\varepsilon_{P}} {16\pi^{2}f_{\ast}^{2}}, \label{eq:alpha_power_generic} \\ |\alpha| &\lesssim \frac{\varepsilon_{\dot f}} {16\pi^{2}f_{\ast}^{2}}.
\label{eq:alpha_chirp_generic}
\end{align}
For reference, at $f_{\ast}=100\,\mathrm{Hz}$ these relations read
\begin{align}
|\alpha| &\lesssim 1.27\times10^{-6}\, \varepsilon_{h}\ \mathrm{s}^{2}, \label{eq:alpha_amplitude_100Hz} \\ |\alpha| &\lesssim 6.33\times10^{-7}\, \varepsilon_{P}\ \mathrm{s}^{2}, \qquad |\alpha| \lesssim 6.33\times10^{-7}\, \varepsilon_{\dot f}\ \mathrm{s}^{2}.
\label{eq:alpha_chirp_100Hz}
\end{align}
These expressions are useful for translating future waveform measurements, but they are not independent LVK posterior bounds. The amplitude correction is correlated with luminosity distance, inclination, detector calibration, and intrinsic source parameters, whereas the chirp correction must be included consistently in a complete inspiral waveform, as we should expect.

A further constraint can be constructed from the accumulated generation phase. Applying the stationary phase approximation to Eq.~\eqref{eq:corrected_chirp_bounds} gives, in the local approximation,
\begin{equation}
\delta\Psi_{\mathrm{gen}}(f) = -\frac{15}{2} \frac{ \pi^{1/3}\alpha }{ (G\mathcal{M}_{c})^{5/3} } f^{1/3}.
\label{eq:generation_phase_alpha}
\end{equation}
Introducing
\begin{equation}
u=(\pi GMf)^{1/3}, \qquad \eta=\frac{m_{1}m_{2}}{M^{2}},
\end{equation}
this correction takes the parameterized post Einsteinian form
\begin{equation}
\delta\Psi_{\mathrm{gen}} = \beta_{\mathrm{ppE}}u, \qquad \beta_{\mathrm{ppE}} = -\frac{15\alpha} {2\eta(GM)^{2}}.
\label{eq:ppE_alpha_mapping}
\end{equation}
It has the frequency scaling of a relative $3$PN phase contribution \cite{Yunes:2009ke}. A posterior on $\beta_{\mathrm{ppE}}$ would therefore translate into
\begin{equation}
|\alpha| \lesssim \frac{2}{15} \eta(GM)^{2} |\beta_{\mathrm{ppE}}|.
\label{eq:alpha_ppE_bound}
\end{equation}
A numerical catalog constraint should not be assigned through Eq.~\eqref{eq:alpha_ppE_bound} without a dedicated analysis. At the same PN order, the standard GR contributions, spin effects, waveform systematics, and possible Ho\v{r}ava--Lifshitz corrections to the conservative binary dynamics must be included simultaneously. Since the mapping also depends on $M$ and $\eta$, a universal bound on $\alpha$ requires a joint event by event or hierarchical inference instead of the direct use of a catalog posterior on a common fractional PN coefficient.

We do not quote a separate mismatch based bound. A template mismatch is obtained only after noise weighting and maximization over time, phase, amplitude, and other source parameters; it cannot in general be identified with $\langle|\Delta h/h|^{2}\rangle$. The direct replacement of a mismatch threshold by a bound on $\alpha$ would consequently neglect important parameter degeneracies.


\subsection{Resulting hierarchy of constraints}
\label{subsec:alpha_constraint_hierarchy}

The current direct observational constraint on the $k^{4}$ Ho\v{r}ava--Lifshitz coefficient is therefore
\begin{equation}
-6.2\times10^{2}\ \mathrm{eV}^{-2} < \alpha < 1.9\times10^{2}\ \mathrm{eV}^{-2} \qquad (90\%~\text{credible interval}),
\label{eq:final_alpha_observational_bound}
\end{equation}
obtained from the GWTC--4.0 modified dispersion posterior. The GW170817/GRB\,170817A speed measurement provides a weaker but independent cross check. The amplitude, luminosity, chirp rate, and generation phase relations derived above define additional channels for a theory specific waveform analysis. At the accuracy allowed by Eq.~\eqref{eq:final_alpha_observational_bound}, their local fractional corrections are negligible in the LVK band, while the propagation phase remains measurable because it accumulates over cosmological distances.


\section{Conclusion}

In this work, we investigated the generation and propagation of gravitational waves in the leading parity--even infrared truncation of Ho\v{r}ava--Lifshitz gravity, characterized by the tensor dispersion relation $\omega^{2}=k^{2}+\alpha k^{4}$. We restricted the analysis to the transverse--traceless sector around Minkowski spacetime and found that the isotropic four--spatial--derivative operator preserved the conventional plus and cross polarizations. Both tensor modes obeyed the same dispersion relation, and no polarization mixing, helicity splitting, gravitational birefringence, or additional tensor mode arose.

We constructed the retarded Green function of the modified wave operator and derived the radiation--zone waveform to first order in $\alpha$. The local derivative expansion remained supported on the general relativistic light cone at finite perturbative order, whereas the exact kernel retained the dispersive propagation associated with the modified frequency. The waveform acquired a frequency--dependent amplitude renormalization and a phase correction that accumulated over the source--observer distance. The phase--resummed representation preserved the overall $1/r$ decay and removed the apparent nondecaying contribution generated by the direct expansion of the propagation phase.

For a binary black hole system in a quasi--circular orbit, we obtained the polarization waveforms for an arbitrary observation direction. The standard quadrupolar angular dependence remained unchanged, while both tensor polarizations acquired the same amplitude factor and propagation phase. The polarization trajectories therefore remained circular, and the transverse deformation retained the usual tensor pattern.

We also derived the angular energy flux, total luminosity, and adiabatic chirp evolution. In terms of the observed gravitational wave frequency $f$, the leading fractional corrections were $\frac{\Delta h_A}{h_A^{\mathrm{GR}}}  =-8\pi^{2}\alpha f^{2}$, and $ \frac{\Delta P}{P_{\mathrm{GR}}}  =\frac{\Delta\dot{f}}{\dot{f}_{\mathrm{GR}}}  =-16\pi^{2}\alpha f^{2}$. The accumulated generation phase displayed the frequency dependence of a relative third post--Newtonian contribution and supplied a source--dynamics channel complementary to the propagation effect.

Finally, we mapped the Ho\v{r}ava--Lifshitz coefficient onto the LIGO--Virgo--KAGRA modified--dispersion parametrization. The GWTC--4.0 posterior yielded $-6.2\times10^{2}\,\mathrm{eV}^{-2} <\alpha< 1.9\times10^{2}\,\mathrm{eV}^{-2}$ at $90\%$ credibility. This interval implied $\ell_{\alpha}=\sqrt{|\alpha|}\lesssim4.91\times10^{-6}\,\mathrm{m}$ and $\Lambda_{\alpha}=|\alpha|^{-1/2}\gtrsim4.02\times10^{-2}\,\mathrm{eV}$. The GW170817/GRB~170817A comparison supplied a weaker but independent propagation--speed constraint. Throughout the LVK frequency band, the condition $|\alpha|(2\pi f)^{2}\ll1$ remained satisfied by many orders of magnitude, while the accumulated propagation phase remained sensitive to the large source distances. In summary, these results showed that the leading parity--even Ho\v{r}ava--Lifshitz correction modified the waveform amplitude, luminosity, chirp evolution, and propagation phase, but left the tensor polarization content unchanged. The parity--odd $k^{5}$ contribution, the ultraviolet $k^{6}$ term, corrections to the conservative binary dynamics, and a dedicated full--waveform inference remained outside the scope of the analysis and were reserved for future investigations.


\section{Acknowledgments}

\hspace{0.5cm}

A. A. Araújo Filho is supported by Conselho Nacional de Desenvolvimento Cient\'{\i}fico e Tecnol\'{o}gico (CNPq), project number 150223/2025-0. N. H. is supported by Conselho Nacional de Desenvolvimento Cient\'{\i}fico e Tecnol\'{o}gico (CNPq), project number 152891/2025-0. N. H. is grateful for the support provided by three COST Actions: CA21106 (COSMIC WISPers in the Dark Universe: Theory, Astrophysics and Experiments), CA21136 (Addressing Observational Tensions in Cosmology with Systematics and Fundamental Physics, also known as CosmoVerse), and CA23130 (Bridging High and Low Energies in Search of Quantum Gravity, or BridgeQG). I. P. L. acknowledges partial financial support from the National Council for Scientific and Technological Development (CNPq), under grant No. 312547/2023-4. In addition, I. P. L. would like to acknowledge the networking support of the COST Actions BridgeQG (CA23130), RQI (CA23115) and FuSe (CA24101), both funded by COST — the European Cooperation in Science and Technology.

\section{Data Availability Statement}

Data Availability Statement: No Data associated in the manuscript


\bibliographystyle{ieeetr}
\bibliography{main}

\begin{thebibliography}{10}

\bibitem{Stelle:1977}
K.~S. Stelle, ``Renormalization of higher-derivative quantum gravity,'' {\em
  Phys. Rev. D}, vol.~16, pp.~953--969, 1977.

\bibitem{Horava:2009uw}
P.~Horava, ``{Quantum Gravity at a Lifshitz Point},'' {\em Phys. Rev. D},
  vol.~79, p.~084008, 2009.

\bibitem{Horava:2009if}
P.~Horava, ``{Spectral Dimension of the Universe in Quantum Gravity at a
  Lifshitz Point},'' {\em Phys. Rev. Lett.}, vol.~102, p.~161301, 2009.

\bibitem{SotiriouVisserWeinfurtner:2009}
T.~P. Sotiriou, M.~Visser, and S.~Weinfurtner, ``Phenomenologically viable
  lorentz-violating quantum gravity,'' {\em Phys. Rev. Lett.}, vol.~102,
  p.~251601, 2009.

\bibitem{Mukohyama:2010review}
S.~Mukohyama, ``Ho\v{r}ava--lifshitz cosmology: A review,'' {\em Class. Quant.
  Grav.}, vol.~27, p.~223101, 2010.

\bibitem{Sotiriou:2011review}
T.~P. Sotiriou, ``Ho\v{r}ava--lifshitz gravity: A status report,'' {\em J.
  Phys. Conf. Ser.}, vol.~283, p.~012034, 2011.

\bibitem{Charmousis:2009}
C.~Charmousis, G.~Niz, A.~Padilla, and P.~M. Saffin, ``Strong coupling in
  ho\v{r}ava gravity,'' {\em JHEP}, vol.~08, p.~070, 2009.

\bibitem{PapazoglouSotiriou:2010}
A.~Papazoglou and T.~P. Sotiriou, ``Strong coupling in extended
  ho\v{r}ava--lifshitz gravity,'' {\em Phys. Lett. B}, vol.~685, pp.~197--200,
  2010.

\bibitem{BlasPujolasSibiryakov:2010}
D.~Blas, O.~Pujol\`as, and S.~Sibiryakov, ``Consistent extension of ho\v{r}ava
  gravity,'' {\em Phys. Rev. Lett.}, vol.~104, p.~181302, 2010.

\bibitem{BlasPujolasSibiryakov:2011}
D.~Blas, O.~Pujol\`as, and S.~Sibiryakov, ``Models of non-relativistic quantum
  gravity: The good, the bad and the healthy,'' {\em JHEP}, vol.~04, p.~018,
  2011.

\bibitem{Jacobson:2010}
T.~Jacobson, ``Extended ho\v{r}ava gravity and einstein--aether theory,'' {\em
  Phys. Rev. D}, vol.~81, p.~101502, 2010.

\bibitem{Barvinsky:2016}
A.~O. Barvinsky, D.~Blas, M.~Herrero-Valea, S.~M. Sibiryakov, and C.~F.
  Steinwachs, ``Renormalization of ho\v{r}ava gravity,'' {\em Phys. Rev. D},
  vol.~93, p.~064022, 2016.

\bibitem{Barvinsky:2017}
A.~O. Barvinsky, D.~Blas, M.~Herrero-Valea, S.~M. Sibiryakov, and C.~F.
  Steinwachs, ``Ho\v{r}ava gravity is asymptotically free in $2+1$
  dimensions,'' {\em Phys. Rev. Lett.}, vol.~119, p.~211301, 2017.

\bibitem{BarvinskyKurovSibiryakov:2024}
A.~O. Barvinsky, A.~V. Kurov, and S.~M. Sibiryakov, ``Renormalization group
  flow of projectable ho\v{r}ava gravity in $(3+1)$ dimensions,'' 2024.

\bibitem{Wang:2017review}
A.~Wang, ``Ho\v{r}ava gravity at a lifshitz point: A progress report,'' {\em
  Int. J. Mod. Phys. D}, vol.~26, p.~1730014, 2017.

\bibitem{HerreroValea:2023}
M.~Herrero-Valea, ``The status of ho\v{r}ava gravity,'' 2023.

\bibitem{Mukohyama:2009}
S.~Mukohyama, ``Scale-invariant cosmological perturbations from
  ho\v{r}ava--lifshitz gravity without inflation,'' {\em JCAP}, vol.~06,
  p.~001, 2009.

\bibitem{TakahashiSoda:2009}
T.~Takahashi and J.~Soda, ``Chiral primordial gravitational waves from a
  lifshitz point,'' {\em Phys. Rev. Lett.}, vol.~102, p.~231301, 2009.

\bibitem{Wang:2010tensor}
A.~Wang, ``Vector and tensor perturbations in ho\v{r}ava--lifshitz cosmology,''
  {\em Phys. Rev. D}, vol.~82, p.~124063, 2010.

\bibitem{WangWuZhaoZhu:2013}
A.~Wang, Q.~Wu, W.~Zhao, and T.~Zhu, ``Polarizing primordial gravitational
  waves by parity violation,'' {\em Phys. Rev. D}, vol.~87, p.~103512, 2013.

\bibitem{LIGO:2016GW150914}
{LIGO Scientific Collaboration and Virgo Collaboration}, ``Observation of
  gravitational waves from a binary black hole merger,'' {\em Phys. Rev.
  Lett.}, vol.~116, p.~061102, 2016.

\bibitem{LIGO:2017GW170817}
{LIGO Scientific Collaboration and Virgo Collaboration}, ``Gw170817:
  Observation of gravitational waves from a binary neutron star inspiral,''
  {\em Phys. Rev. Lett.}, vol.~119, p.~161101, 2017.

\bibitem{LIGO:2017MultiMessenger}
{LIGO Scientific Collaboration and Virgo Collaboration and Fermi GBM and
  INTEGRAL}, ``Gravitational waves and gamma-rays from a binary neutron star
  merger: Gw170817 and grb 170817a,'' {\em Astrophys. J. Lett.}, vol.~848,
  p.~L13, 2017.

\bibitem{LVK:2026GWTC5}
{LIGO Scientific Collaboration and Virgo Collaboration and KAGRA
  Collaboration}, ``Gwtc--5.0: Observations from the second part of the fourth
  ligo--virgo--kagra observing run and updates to the gravitational-wave
  transient catalog,'' 2026.

\bibitem{LVK:2026GWTC4TestsII}
{LIGO Scientific Collaboration and Virgo Collaboration and KAGRA
  Collaboration}, ``Gwtc--4.0: Tests of general relativity. ii. parameterized
  tests,'' 2026.

\bibitem{md1}
R.~C. Myers and M.~Pospelov, ``Ultraviolet modifications of dispersion
  relations in effective field theory,'' {\em Physical Review Letters},
  vol.~90, no.~21, p.~211601, 2003.

\bibitem{md2}
B.~R. Majhi and E.~C. Vagenas, ``Modified dispersion relation, photon`s
  velocity, and unruh effect,'' {\em Physics Letters B}, vol.~725, no.~4-5,
  pp.~477--480, 2013.

\bibitem{md3}
F.~Girelli, S.~Liberati, and L.~Sindoni, ``Planck-scale modified dispersion
  relations and finsler geometry,'' {\em Physical Review D—Particles, Fields,
  Gravitation, and Cosmology}, vol.~75, no.~6, p.~064015, 2007.

\bibitem{md4}
G.~Rosati, G.~Amelino-Camelia, A.~Marciano, and M.~Matassa,
  ``Planck-scale-modified dispersion relations in frw spacetime,'' {\em
  Physical Review D}, vol.~92, no.~12, p.~124042, 2015.

\bibitem{md5}
F.~Girelli, S.~Liberati, R.~Percacci, and C.~Rahmede, ``Modified dispersion
  relations from the renormalization group of gravity,'' {\em Classical and
  Quantum Gravity}, vol.~24, no.~16, pp.~3995--4008, 2007.

\bibitem{md6}
Y.~Ling, B.~Hu, and X.~Li, ``Modified dispersion relations and black hole
  physics,'' {\em Physical Review D—Particles, Fields, Gravitation, and
  Cosmology}, vol.~73, no.~8, p.~087702, 2006.

\bibitem{md7}
A.~Sefiedgar, K.~Nozari, and H.~R. Sepangi, ``Modified dispersion relations in
  extra dimensions,'' {\em Physics Letters B}, vol.~696, no.~1-2, pp.~119--123,
  2011.

\bibitem{md8}
A.~A. Ara{\'u}jo~Filho, ``{Particle production induced by a Lorentzian
  non-commutative spacetime},'' {\em Annals Phys.}, vol.~481, p.~170167, 2025.

\bibitem{md9}
J.~Furtado, H.~Hassanabadi, J.~A. A.~S. Reis, {\em et~al.}, ``Thermal analysis
  of photon-like particles in rainbow gravity,'' {\em arXiv preprint
  arXiv:2305.08587}, 2023.

\bibitem{md10}
A.~A. Ara{\'u}jo~Filho, ``Particles in loop quantum gravity formalism: a
  thermodynamical description,'' {\em Annalen der Physik}, vol.~534, no.~12,
  p.~2200383, 2022.

\bibitem{md11}
A.~A. Ara{\'u}jo~Filho, J.~Furtado, J.~A. A.~S. Reis, and J.~Silva,
  ``Thermodynamical properties of an ideal gas in a traversable wormhole,''
  {\em Classical and Quantum Gravity}, vol.~40, no.~24, p.~245001, 2023.

\bibitem{md12}
A.~A. Ara{\'u}jo~Filho and J.~A. A.~S. Reis, ``How does geometry affect quantum
  gases?,'' {\em International Journal of Modern Physics A}, vol.~37,
  no.~11n12, p.~2250071, 2022.

\bibitem{md13}
A.~A. Ara{\'u}jo~Filho, {\em Thermal aspects of field theories}.
\newblock Amazon. com, 2022.

\bibitem{md14}
A.~A. Ara{\'u}jo~Filho, ``Implications of a simpson--visser solution in
  verlinde’s framework,'' {\em The European Physical Journal C}, vol.~84,
  no.~1, p.~73, 2024.

\bibitem{md15}
M.~A. Anacleto, J.~A.~V. Campos, F.~A. Brito, E.~Maciel, and E.~Passos,
  ``{Scattering and absorption by extra-dimensional black holes with GUP},''
  {\em Nucl. Phys. B}, vol.~1006, p.~116617, 2024.

\bibitem{md16}
M.~A. Anacleto, J.~A.~V. Campos, F.~A. Brito, and E.~Passos, ``{Quasinormal
  modes and shadow of a Schwarzschild black hole with GUP},'' {\em Annals
  Phys.}, vol.~434, p.~168662, 2021.

\bibitem{md17}
M.~A. Anacleto, F.~A. Brito, S.~S. Cruz, and E.~Passos, ``{Noncommutative
  correction to the entropy of Schwarzschild black hole with GUP},'' {\em Int.
  J. Mod. Phys. A}, vol.~36, no.~03, p.~2150028, 2021.

\bibitem{md18}
M.~A. Anacleto, F.~A. Brito, B.~R. Carvalho, and E.~Passos, ``{Noncommutative
  correction to the entropy of BTZ black hole with GUP},'' {\em Adv. High
  Energy Phys.}, vol.~2021, p.~6633684, 2021.

\bibitem{md19}
M.~A. Anacleto, F.~A. Brito, J.~A.~V. Campos, and E.~Passos,
  ``{Quantum-corrected scattering and absorption of a Schwarzschild black hole
  with GUP},'' {\em Phys. Lett. B}, vol.~810, p.~135830, 2020.

\bibitem{MirshekariYunesWill:2012}
S.~Mirshekari, N.~Yunes, and C.~M. Will, ``Constraining lorentz-violating,
  modified dispersion relations with gravitational waves,'' {\em Phys. Rev. D},
  vol.~85, p.~024041, 2012.

\bibitem{KosteleckyMewes:2016}
V.~A. Kosteleck\'y and M.~Mewes, ``Testing local lorentz invariance with
  gravitational waves,'' {\em Phys. Lett. B}, vol.~757, pp.~510--514, 2016.

\bibitem{EzquiagaEtAl:2022}
J.~M. Ezquiaga, W.~Hu, M.~Lagos, M.-X. Lin, and F.~Xu, ``Modified
  gravitational-wave propagation with higher modes and its degeneracies with
  lensing,'' {\em JCAP}, vol.~08, p.~016, 2022.

\bibitem{BakaEtAl:2025}
T.~Baka, B.~Cirok, K.~Haris, J.~Noller, and N.~V. Krishnendu, ``Testing general
  relativity with gravitational waves: Improving and extending modified
  dispersion relation tests,'' 2025.

\bibitem{GongEtAl:2022}
C.~Gong, T.~Zhu, R.~Niu, Q.~Wu, J.-L. Cui, X.~Zhang, W.~Zhao, and A.~Wang,
  ``Gravitational-wave constraints on lorentz and parity violations in gravity:
  High-order spatial-derivative cases,'' {\em Phys. Rev. D}, vol.~105,
  p.~044034, 2022.

\bibitem{WangYanZhuZhao:2025}
Q.~Wang, J.-M. Yan, T.~Zhu, and W.~Zhao, ``Modified gravitational-wave
  propagations in linearized gravity with lorentz and diffeomorphism violations
  and their gravitational-wave constraints,'' 2025.

\bibitem{AraujoFilhoHeidariLobo:2026}
A.~A. Ara{\'u}jo~Filho, N.~Heidari, and I.~P. Lobo, ``{Propagation effects of
  Lorentz violation in gravitational waves},'' {\em Eur. Phys. J. C}, vol.~86,
  no.~6, p.~630, 2026.

\bibitem{A1}
A.~A. Ara{\'u}jo~Filho, N.~Heidari, and I.~P. Lobo, ``{Gravitational waves in a
  minimal gravitational SME},'' {\em Phys. Lett. B}, vol.~875, p.~140350, 2026.

\bibitem{A2}
A.~A. Ara{\'u}jo~Filho, ``{Gravitational waves in metric{\textendash}affine
  bumblebee gravity},'' {\em Eur. Phys. J. C}, vol.~86, no.~5, p.~578, 2026.

\bibitem{A3}
K.~M. Amarilo, M.~B.~F. Filho, A.~A.~A. Filho, and J.~A. A.~S. Reis,
  ``{Gravitational waves effects in a Lorentz{\textendash}violating
  scenario},'' {\em Phys. Lett. B}, vol.~855, p.~138785, 2024.

\bibitem{BlasSanctuary:2011}
D.~Blas and H.~Sanctuary, ``Gravitational radiation in ho\v{r}ava gravity,''
  {\em Phys. Rev. D}, vol.~84, p.~064004, 2011.

\bibitem{GongHouPapantonopoulosTzortzis:2018}
Y.~Gong, S.~Hou, E.~Papantonopoulos, and D.~Tzortzis, ``Gravitational waves and
  the polarizations in ho\v{r}ava gravity after gw170817,'' {\em Phys. Rev. D},
  vol.~98, p.~104017, 2018.

\bibitem{Maggiore:2007ulw}
M.~Maggiore, {\em {Gravitational Waves. Vol. 1: Theory and Experiments}}.
\newblock Oxford University Press, 2007.

\bibitem{Goldstein:2002}
H.~Goldstein, C.~Poole, and J.~Safko, {\em {Classical Mechanics}}.
\newblock San Francisco, USA: Addison-Wesley, 3rd~ed., 2002.

\bibitem{LVK:GWTC5}
{The LIGO Scientific Collaboration}, {the Virgo Collaboration}, and {the KAGRA
  Collaboration}, ``{GWTC-5.0: Observations from the Second Part of the Fourth
  LIGO--Virgo--KAGRA Observing Run and Updates to the Gravitational-Wave
  Transient Catalog},'' 2026.

\bibitem{LVK:GWTC4TestsII}
{The LIGO Scientific Collaboration}, {the Virgo Collaboration}, and {the KAGRA
  Collaboration}, ``{GWTC-4.0: Tests of General Relativity. II. Parameterized
  Tests},'' 2026.

\bibitem{Mirshekari:2011yq}
S.~Mirshekari, N.~Yunes, and C.~M. Will, ``{Constraining Lorentz-Violating,
  Modified Dispersion Relations with Gravitational Waves},'' {\em Phys. Rev.
  D}, vol.~85, p.~024041, 2012.

\bibitem{Ezquiaga:2022MDR}
J.~M. Ezquiaga, W.~Hu, M.~Lagos, M.-X. Lin, and F.~Xu, ``{Modified
  Gravitational Wave Propagation with Higher Modes and Its Degeneracies with
  Lensing},'' {\em JCAP}, vol.~08, p.~016, 2022.

\bibitem{Baka:2025MDR}
T.~Baka, B.~Cirok, K.~Haris, J.~Noller, and N.~V. Krishnendu, ``{Testing
  General Relativity with Gravitational Waves: Improving and Extending Modified
  Dispersion Relation Tests},'' 2025.

\bibitem{LIGOScientific:2017zic}
{LIGO Scientific Collaboration}, {Virgo Collaboration}, {Fermi GBM}, and
  {INTEGRAL}, ``{Gravitational Waves and Gamma-Rays from a Binary Neutron Star
  Merger: GW170817 and GRB 170817A},'' {\em Astrophys. J. Lett.}, vol.~848,
  no.~2, p.~L13, 2017.

\bibitem{Yunes:2009ke}
N.~Yunes and F.~Pretorius, ``{Fundamental Theoretical Bias in Gravitational
  Wave Astrophysics and the Parameterized Post-Einsteinian Framework},'' {\em
  Phys. Rev. D}, vol.~80, p.~122003, 2009.

\end{thebibliography}

\end{document}